\documentclass[%
 reprint,
 amsmath,
 amssymb,
 aps,
 prl,
]{revtex4-1}

\usepackage{graphicx}
\usepackage{dcolumn}
\usepackage{bm}
\usepackage{physics}
\usepackage{hyperref}
\usepackage{amsmath}
\usepackage{multirow}
\DeclareMathOperator*{\argmax}{argmax}
\newcommand\numberthis{\addtocounter{equation}{1}\tag{\theequation}}
\makeatletter
\newcommand{\xRightarrow}[2][]{\ext@arrow 0359\Rightarrowfill@{#1}{#2}}
\makeatother

\begin{document}

\preprint{APS/123-QED}

\title{Preparation of Metrological States in Dipolar-Interacting Spin Systems}

\author{Tian-Xing Zheng\textsuperscript{1,2}}
\author{Anran Li\textsuperscript{1}}
\author{Jude Rosen\textsuperscript{1}}
\author{Sisi Zhou\textsuperscript{1,3}}
\author{Martin Koppenh\"ofer\textsuperscript{1}} 
\author{Ziqi Ma\textsuperscript{4,5}}
\author{Frederic T. Chong\textsuperscript{4}} 
\author{Aashish A. Clerk\textsuperscript{1}}
\author{Liang Jiang\textsuperscript{1}} 
\author{Peter C. Maurer\textsuperscript{1}}

\affiliation{\textsuperscript{1}Pritzker School of Molecular Engineering, University of Chicago, Chicago, Illinois 60637, USA}
\affiliation{\textsuperscript{2}Department of Physics, University of Chicago, Chicago, Illinois 60637, USA}
\affiliation{\textsuperscript{3}Institute for Quantum Information and Matter, California Institute of Technology, Pasadena, California 91125, USA}
\affiliation{\textsuperscript{4}Department of Computer Science, University of Chicago, Chicago, Illinois 60637, USA}
\affiliation{\textsuperscript{5}Microsoft, Redmond, Washington 98052, USA}

\begin{abstract}
Spin systems are an attractive candidate for quantum-enhanced metrology. Here we develop a variational method to generate metrological states in small dipolar-interacting ensembles with limited qubit controls and unknown spin locations. The generated states enable sensing beyond the standard quantum limit (SQL) and approaching the Heisenberg limit (HL). Depending on the circuit depth and the level of readout noise, the resulting states resemble Greenberger-Horne-Zeilinger (GHZ) states or Spin Squeezed States (SSS). Sensing beyond the SQL holds in the presence of finite spin polarization and a non-Markovian noise environment. 

\end{abstract}

\maketitle
\paragraph{Introduction.}
Spin systems have emerged as a promising platform for quantum sensing [\onlinecite{degen2017quantum,barry2020sensitivity,schirhagl2014nitrogen,tetienne2021quantum}] with applications ranging from tests of fundamental physics [\onlinecite{hensen2015loophole},\onlinecite{marti2018imaging}] to mapping fields and temperature profiles in condensed matter systems and life sciences [\onlinecite{schirhagl2014nitrogen}]. Improving the sensitivity of these qubit sensors has so far largely relied on increasing the number of sensing spins and extending spin coherence through material engineering and coherent control. However, with increasing spin density, dipolar interactions between individual sensor spins cause single-qubit dephasing [\onlinecite{yan2013observation},\onlinecite{childress2006coherent}] and, in the absence of advanced dynamical decoupling [\onlinecite{waugh1968approach,mehring2012principles,choi2020robust}], set a limit to the sensitivity.

Although dipolar interactions in dense spin ensembles lead to complex evolution, they can provide a resource for the creation of metrological states that enable sensing beyond the SQL. Current approaches to create such states (i.e., GHZ states and SSS) either require all-to-all interactions [\onlinecite{pedrozo2020entanglement,kitagawa1993squeezed,bilitewski2021dynamical}] or single-qubit addressability [\onlinecite{chen2020parallel},\onlinecite{neumann2008multipartite}], which are challenging to implement experimentally. An alternative approach that relies on adiabatic state preparation requires less control but results in preparation times that increase exponentially with system size [\onlinecite{cappellaro2009quantum},\onlinecite{choi2017quantum}], leaving this method susceptible to dephasing.

\begin{figure}[h]
\includegraphics[width=86mm]{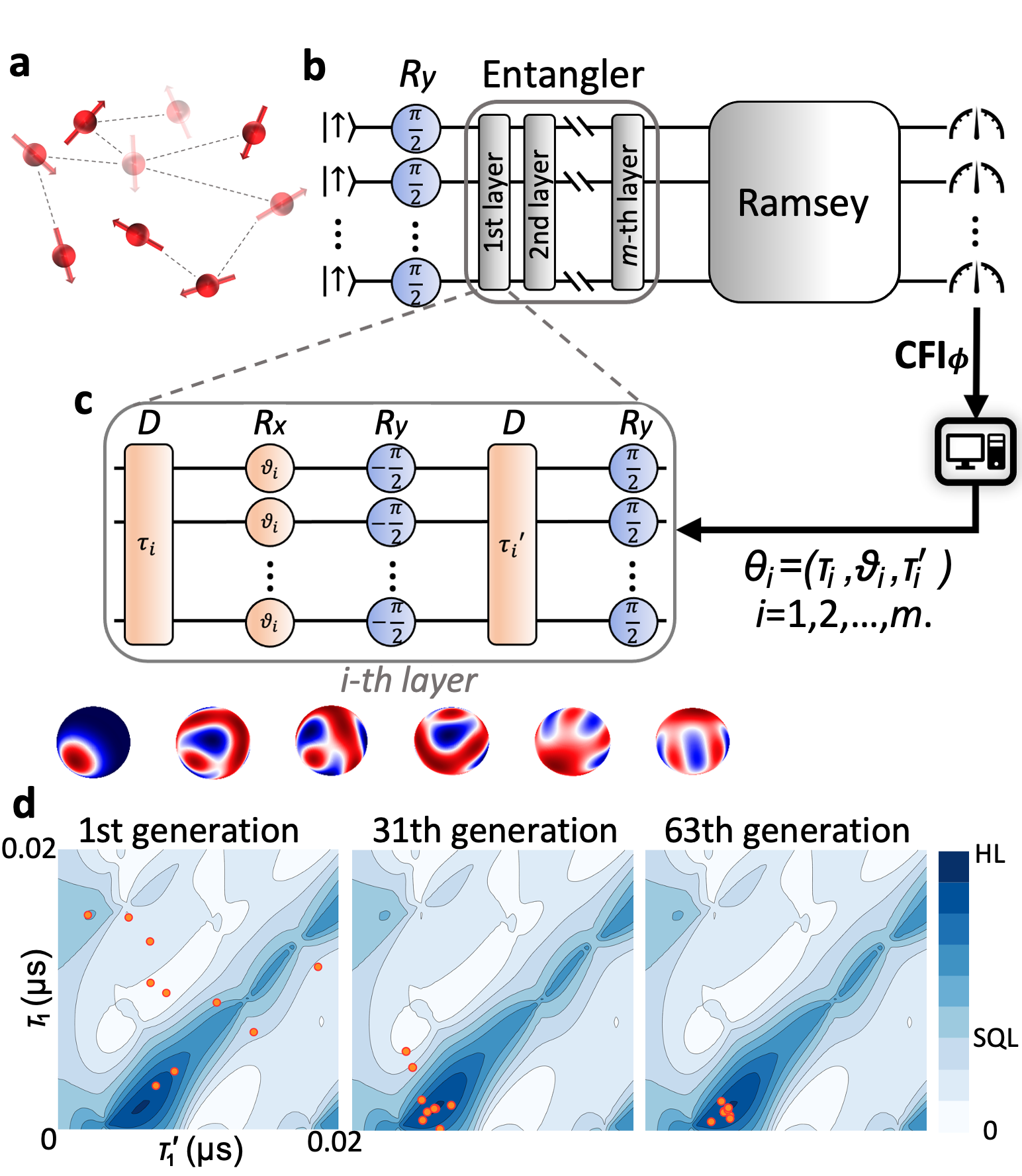}
\caption{\label{fig:1}
(a) Schematic of a dipolar-interacting spin ensemble in a 3D-random configuration. (b) The quantum circuit consists of three parts: a sequence for generating entanglement (entangler), phase accumulation (Ramsey) and single-qubit readout in the $S_z$ basis. Dipolar interactions during Ramsey interference are eliminated by dynamical decoupling [\onlinecite{yan2013observation},\onlinecite{waugh1968approach},\onlinecite{zhou2020quantum}]. The measurement outcome is processed on a classical computer and used to determine the next generation for $\bm{\theta}$. (c) Gate sequence of each variational layer and the Wigner distributions for a 5-spin state after each gate. (d) Illustration of an optimization process on a 3-spin system with $m=1$. The contour plots show the 2D projection of the multidimensional $\bm{\theta}$ space for fixed $\vartheta_1$. The orange points mark the sampling positions in the parameter space. Convergence to the global maximum is reached in the 63th generation. 
}
\end{figure}

Variational methods provide a powerful tool for controlling many-body quantum systems [\onlinecite{cerezo2021variational,kokail2019self,li2017efficient,leozhou2020QAOA}]. 
Such methods have been proposed for Rydberg-interacting atomic systems [\onlinecite{kaubruegger2019variational},\onlinecite{kaubruegger2021quantum}] and demonstrated in trapped ions [\onlinecite{marciniak2021optimal}].
However, these techniques rely on effective all-to-all interactions (e.g. almost constant interaction strength inside the Rydberg radius [\onlinecite{kaubruegger2019variational},\onlinecite{borish2020transverse},\onlinecite{bernien2017probing}]) which are generally absent in solid-state spin ensembles.
In this work, we develop a variational algorithm that drives dipolar-interacting spin systems [Fig.~\ref{fig:1}(a)] into highly entangled states.
The resulting states can be subsequently used for Ramsey-interferometry-based single parameter estimation [\onlinecite{degen2017quantum}]. The required system control relies solely on uniform single-qubit rotations and free evolution under dipolar interactions. 
The optimization can be directly performed on an experimental device using only its measurement outcomes without the need to know the spatial distribution of the spins (later referred as `spin configuration').
Potential experimental platforms include dipolar-interacting ensembles of NV centers, nitrogen defects in diamond (P1), rare-earth-doped crystals, and ultra-cold molecules.

\paragraph{Variational Ansatz.}
As shown in Fig.~\ref{fig:1}(b), the variational circuit $\mathcal{S}(\bm{\theta}) = \mathcal{U}_m ... \mathcal{U}_2\mathcal{U}_1$ is constructed by $m$ layers of unitary operations. Each $\mathcal{U}_i$ consists of the parameterized control gates
\begin{equation}
    \mathcal{U}_i = R_y\left(\frac{\pi}{2}\right)D\left(\tau'_i\right)R_y\left(-\frac{\pi}{2}\right)R_x\left(\vartheta_i\right)D\left(\tau_i\right),
    \label{eq:1}
\end{equation}
where $R_{\mu}(\vartheta) =  \exp( - i \vartheta \sum_{j=1}^N S_{j}^{\mu})$ are single-qubit rotations and $S_j^\mu$ $(\mu\in\{x,y,z\})$ is the $\mu$ component of the $j$-th spin operator. $D(\tau) = \exp(-i\tau H_\text{dd}/\hbar)$ is the time evolution operator of the spin ensemble under dipolar-interaction Hamiltonian $H_\text{dd} = \sum_{i<j} V_{ij}  (2 S^z_i S^z_j - S^x_i S^x_j- S^y_i S^y_j)$. The coupling strength between two spins at positions $\bm{r}_i$ and $\bm{r}_j$ is
\begin{equation}
    V_{ij} = \frac{\mu_0}{4\pi}\frac{\gamma_i\gamma_j \hbar^2}{\left|\bm{r}_i - \bm{r}_j\right|^3}\frac{(1 - 3\cos\beta_{ij})}{2},
    \label{eq:2}
\end{equation}
with $\mu_0$ the vacuum permeability, $\hbar$ the reduced Planck constant, $\gamma$ the spin's gyromagnetic ratio, and $\beta_{ij}$ the angle between the line segment connecting ($\bm{r}_i$, $\bm{r}_j$) and the direction of bias magnetic field.
An evolutionary algorithm [\onlinecite{hansen2016cma},\onlinecite{SM}] is applied on the $m$-layer circuit which contains $3m$ free parameters constituting the vector $\bm{\theta} = (\tau_1,\vartheta_1,\tau'_1,...,\tau_i,\vartheta_i,\tau'_i,...,\tau_m,\vartheta_m,\tau'_m)$.
Each $\tau_i$ is restricted to $\tau_i \in [0, 1/\bar{f}_\text{dd}]$ where $\bar{f}_\text{dd}$ is the average nearest-neighbor interaction strength for the considered spin configuration.
The Ansatz in Eq.~(\ref{eq:1}) is the most general set of global single-qubit gates that preserves the initial collective spin direction $\langle\sum_{i}\bm{S}_i\rangle/|\langle\sum_{i}\bm{S}_i\rangle|$, here chosen to be the $x$-direction [\onlinecite{kaubruegger2019variational},\onlinecite{SM}]. Although this Ansatz does not enable universal system control [\onlinecite{schirmer2001complete, albertini2002lie, albertini2021SubspaceControllability,SM}], we show that with increasing circuit depth, sensing near the HL can be achieved. 

\paragraph{Metrological cost function.}
The Ramsey protocol shown in Fig.~\ref{fig:1}(b) encodes the quantity of interest in the accumulated phase $\phi = \omega t_\text{R}$, with $\omega$ the detuning frequency and $t_\text{R}$ the Ramsey sensing time. The Classical Fisher Information (CFI) [\onlinecite{degen2017quantum},\onlinecite{SM}] quantifies how precisely one can estimate an unknown parameter $\phi$ under a measurement basis. Our variational approach treats the spin systems as a black-box for which the algorithm finds a control sequence that maximizes the CFI associated with the parameter estimation problem 
\begin{equation}
    \mathrm{CFI}_{\phi} = \sum_{z} \Tr[P_{z} \rho_{\phi}]  \left(\frac{\partial \log \Tr[P_{z} \rho_{\phi}]}{\partial\phi}\right)^2. 
    \label{eq:5}
 \end{equation}
The sum runs over the $2^N$ basis states $\ket{z} = \otimes_{i=1}^N \ket{s_i^z}$, where $s_i^z$ are the eigenvalues of $S_i^z$. $P_{z} \equiv \ket{z} \bra{z}$ denotes the corresponding measurement operator and $\rho_{\phi}$ the density matrix. 
The CFI is chosen as cost function because it is a measure for the maximal achievable sensitivity for a given measurement basis [\onlinecite{degen2017quantum},\onlinecite{kobayashi2011probability}].
Likewise, $P_z$ provides the maximal information that can be gained from single-qubit measurements. 
However, measurement operators such as parity or total spin polarization result in a smaller outcome space and are therefore more efficient in experimental implementations. While this study optimizes the measurement for $P_z$, the obtained results likewise hold for parity and total spin polarization [\onlinecite{SM}].

\begin{figure}[b]
\includegraphics[width=86mm]{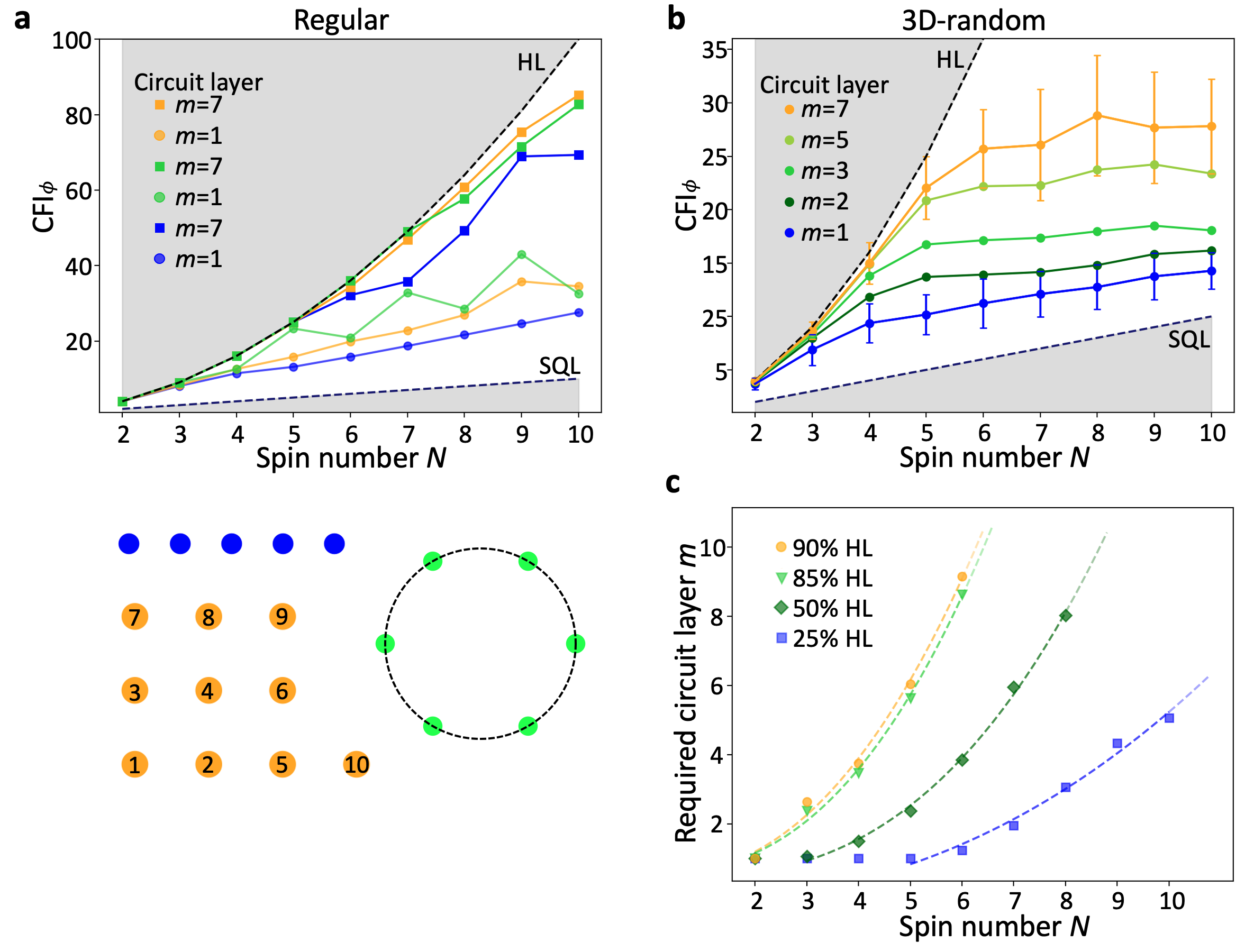}
\caption{\label{fig:2} 
(a) Top: CFI for $m=1$ (circles) and $m=7$ (squares) circuits.  The colors correspond to the configurations shown on the left. Bottom: schematics of different spin configurations. The numbers in the 2D square lattice pattern label the order in which spins are added to form a lattice of size $N$. 
(b) Average CFI for 50 configurations of 3D-randomly distributed spins. (c) Average number of layers required to achieve a CFI within a given percentage of the HL in the case of 3D-random configuration. The fit $m = a N^b + c$ with $b = 2.45$ (goodness of fit $R^2 = 0.996$) serve as a guide to the eye. The same data also fits to an exponential model with slightly lower $R^2=0.995$. 
}
\end{figure}

\begin{figure*}[t]
\includegraphics[width=165mm]{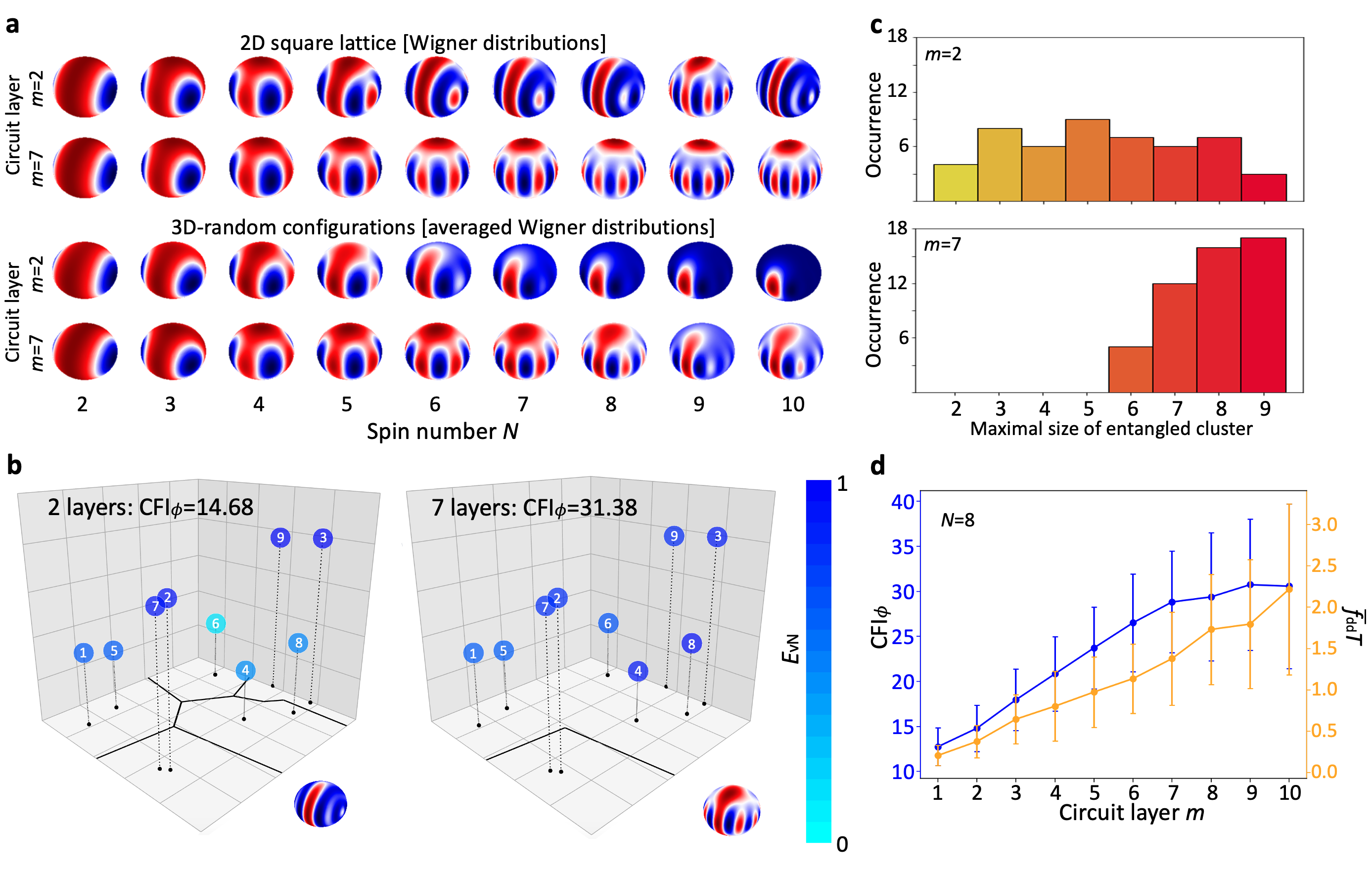}
\caption{\label{fig:3} 
(a) Wigner distributions versus spin number for $m=2$ and $m=7$ in the case of 2D square lattice and 3D-random configurations.
(b) von-Neumann entanglement entropy for one specific 3D-random configuration of 9 spins for $m=2$ and $m=7$. Individual spins are colored according to their von-Neumann entropies. Entangled clusters are marked by solid black lines. 
(c) Histograms depicting the maximal size of entangled clusters for 50 3D-random configurations. 
(d) Average CFI (blue) and state preparation time (orange) versus $m$. The state preparation time is given as a unitless quantity $\bar{f}_\text{dd} T$ with $T = \sum^{m}_{i=1} (\tau_i+\tau'_i)$. 
}
\end{figure*}

\paragraph{Numerical results for regular and disordered spin configurations.}
We start by testing our approach for three distinct regular spin configurations. Figure~\ref{fig:2}(a) shows the CFI after optimization for spins arranged on a linear chain (blue), a two-dimensional (2D) square lattice (orange), and a circle (green). All three configurations result in states with CFI above the SQL. When multiple circuit layers are added, the CFI further improves. Next, we simulate the case of disordered three-dimensional (3D) spin configurations (later referred as 3D-random). In our simulations the spins are randomly located in a box of length $L \propto N^{1/3}$ (constant spin density). Compared to the regular spin array, the disordered case shows a noticeable saturation of the CFI as a funciton of $N$. With increased circuit depth, sensing precision beyond the SQL is maintained. The required circuit depth increases drastically with $N$ [Fig.~\ref{fig:2}(c)].

\paragraph{Characterization of entanglement.}
We investigate the $N$-qubit entangled states created by our variational method. Figure~\ref{fig:3}(a) shows the corresponding Wigner distributions [\onlinecite{hillery1984distribution,dowling1994wigner,koczor2020fast}] for a regular 2D spin array (top) and the average Wigner distributions for 50 different 3D-random spin configurations (bottom). In both cases, the optimized states resemble GHZ states when $N$ is small and $m$ is large. For large $N$ and small $m$, the states are close to SSS. Non-Gaussian states that provide sensitivity beyond the SSS but lower than GHZ states are also generated. Our algorithm tends to drive the systems into a GHZ state, as it has the unique property of attaining the HL in Ramsey spectroscopy [\onlinecite{giovannetti2006quantum}].

For quantitatively analysing the buildup of entanglement, the von-Neumann entanglement entropy ($ E_\text{vN} = -\Tr(\rho_{\text{s}}\log_2\rho_{\text{s}})$) [\onlinecite{nielsen_chuang_2010}] is used as a measure for the degree of entanglement between a spin subsystem ($\rho_{\text{s}}=\Tr_{\text{s}}{\rho_{\text{tot}}}$) and the remaining system. As an example, we explore one case of a 3D-random configuration of 9 spins. Figure~\ref{fig:3}(b) shows the von-Neumann entropy of each spin after employing a 2-layer circuit (left) and a 7-layer circuit (right). In the case of $m=2$, the achieved degree of entanglement is modest with spin No.6 for example showing no substantial entanglement with the remaining spins. When the circuit depth is increased to 7, all spins display substantial entanglement. While the single-particle entropy detects spins unentangled with the remaining system, it does not determine whether all spins are entangled with each other or entanglement is local. We distinguish these two scenarios by identifying the smallest clusters with $E_{\text{vN}}\leq 0.4$. For $m=2$, the spin ensemble segments into 5 clusters [Fig.~\ref{fig:3}(b)], while for $m=7$ only 2 clusters are found. The results verify that multiple layers are required to overcome the anisotropy of the dipolar interaction (Eq.~(\ref{eq:2})) when building up entanglement over the entire system. Finally, in Fig.~\ref{fig:3}(c) we analyze the size of the largest cluster for each of the 50 spin configurations and observe an overall increase of the largest cluster size and a decrease of the variance.

\paragraph{State preparation time.}

Minimizing the preparation time is central in practical applications, as it increases bandwidth, reduces decoherence, and enables more measurement repetitions [\onlinecite{degen2017quantum}]. Figure~\ref{fig:3}(d) shows the average state preparation time for 8 spins in 50 different 3D-random configurations as a function of layer number. The preparation time increases with the layer number and is inversely proportional to the average dipole coupling strength of the nearest-neighbour spins $\bar{f}_\text{dd}$. Compared to adiabatic methods [\onlinecite{cappellaro2009quantum}], our approach results in an $11\times$ reduction of the preparation time to reach the same CFI for identical spin number and density [\onlinecite{SM}].
\begin{figure}[t]
\includegraphics[width=86mm]{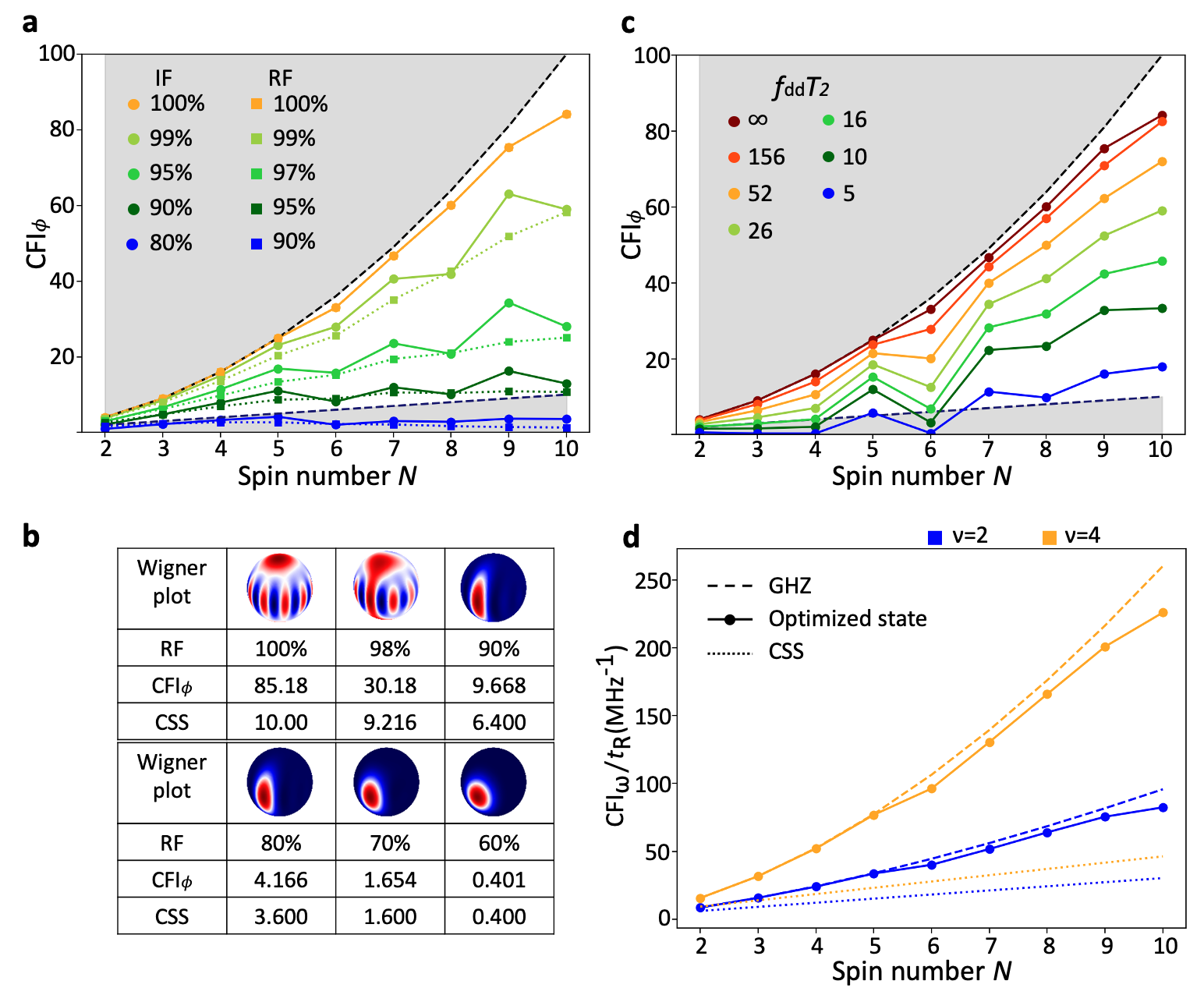}
\caption{\label{fig:4} 
(a) $\text{CFI}_{\phi}$ under finite initialization fidelity (IF) and readout fidelity (RF). Circuit depth $m=5$. $\text{IF}=\frac{N_{\uparrow}-N_{\downarrow}}{N}$, where $N_{\uparrow}$ ($N_{\downarrow}$) denotes the number of spins in the $\ket{\uparrow}$ ($\ket{\downarrow}$) state at the beginning of the sensing protocol in Fig.~\ref{fig:1}(b). $\text{RF} = 1 - p(\downarrow|\uparrow) = 1 - p(\uparrow|\downarrow)$. (b) Optimized Wigner functions and  $\text{CFI}_{\phi}$ of a 10-spin state versus different readout fidelities (RF). For comparison, the row `CSS' represents the  $\text{CFI}_{\phi}$ for a coherent spin state given the same RF. (c) $\text{CFI}_{\phi}$ in the presence of decoherence in the entangler. (d) Ramsey protocol's results of the generated states when considering non-Markovian noise during signal accumulation. Data correspond to the optimized states from 2D square lattice configuration.
}
\end{figure}

\paragraph{State preparation under decoherence, initialization and readout errors}

Until now our analysis assumed full coherence and perfect spin initialization and readout. However, dephasing, initializaiton, and readout errors will be limiting factors in experimental implementations. We next examine the impact of such imperfections on state preparation and sensing. Figure~\ref{fig:4}(a) shows the CFI in the presence of imperfect initialization and finite readout fidelity for spins on a 2D square lattice. For $N\leq10$, beyond-SQL precision is reached with 90\% initialization and 95\% readout fidelity, respectively.

Next, we investigate the resulting states when readout errors are added into the optimizer. Figure~\ref{fig:4}(b) indicates that without readout errors, the Wigner distribution of the resulting state is close to a GHZ state. However, with a finite readout error rate, our algorithm drives the system into a state resembling a SSS. When the readout noise is further increased, the SSS transforms into a coherent spin state (CSS). The results agree with the fact that GHZ states are sensitive to single-spin readout errors while SSS are more robust [\onlinecite{davis2017advantages}]. 

\begin{table*}
\caption{\label{tab:table1}%
Experimental platforms' relative parameters}
\begin{ruledtabular}
\begin{tabular}{c|lllllll}
Systems & $\bar{f}_\text{dd}$ & $T_2^{\text{(DD)}}$ & $\bar{f}_\text{dd} T_2^{\text{(DD)}}$ & $P_\text{ini}$ & $F_\text{readout}$ & $\nu$ \\
\hline
NV ensemble    & $35$kHz  [\onlinecite{zhou2020quantum}] &$7.9(2)\mu$s [\onlinecite{zhou2020quantum}]     & 0.28 & $97.5\%$ [\onlinecite{shields2015efficient}] & $97.5\%$  [\onlinecite{shields2015efficient}] & $2-4$ [\onlinecite{childress2006coherent}]  \\
P1 centers    & $0.92$MHz [\onlinecite{zu2021emergent}] &  $4.4\mu$s [\onlinecite{zu2021emergent}] & 4.0 & $95\%$ [\onlinecite{degen2021entanglement}]& $95\%$ [\onlinecite{degen2021entanglement}]& ?  \\
Rare-Earth crystals & $1.96$MHz [\onlinecite{merkel2021dynamical}] & $2.5\mu$s [\onlinecite{merkel2021dynamical}] & 4.9  & $97\%$ [\onlinecite{chen2020parallel}] & $94.6\%$ [\onlinecite{raha2020optical}] & $2.4\pm0.1$ [\onlinecite{dantec2021twenty}] \\
Cold Molecules & $52$Hz [\onlinecite{yan2013observation}] & $80$ms [\onlinecite{yan2013observation}]  & 4.16  & $97\%$ [\onlinecite{cheuk2020observation}] & $97\%$ [\onlinecite{cheuk2020observation}] & ?  \\
\end{tabular}
\end{ruledtabular}
\end{table*}

During the state preparation, decoherence ($T_2$) reduces entanglement. We assume independent, Markovian dephasing of each spin as described by a Lindblad master equation [\onlinecite{nielsen_chuang_2010}]. Figure~\ref{fig:4}(c) shows the CFI for various $T_2$ times using the previously optimized gate parameters for 2D square lattice. While a finite $T_2$ decreases the CFI, coherence times exceeding $5 / f_\text{dd}$ result in states with beyond-SQL sensitivity. Here, $f_\text{dd}$ denotes the nearest-neighbor interaction strength for 2D square lattice. 
Since performing optimization with imperfections is numerically expensive, the results in Fig.~\ref{fig:4}(a), (c), (d) are obtained by optimizing the parameters in the absence of imperfections and using those parameters to compute the CFI under imperfect conditions. Thus, better results are expected if the optimization is directly run on experiments.

\paragraph{Sensitivity in a non-Markovian environment.}

In addition to impacts on state preparation, dephasing affects performance in Ramsey interferometry. In the presence of spatially uncorrelated Markovian noise, entanglement does not lead to a beyond-SQL scaling [\onlinecite{demkowicz2012elusive},\onlinecite{escher2011general}]. In a non-Markovian environment, such as that of a solid-state spin system, this limitation does not hold [\onlinecite{chin2012quantum},\onlinecite{smirne2016ultimate}]. We examine the performance of our optimized states in a non-Markovian noise environment. We adopt a noise model [\onlinecite{chin2012quantum}] in which the amplitude of single-spin coherence reduces according to
\begin{equation}
    \rho_{01}(t) = \rho_{01}(0) e^{-\left(\frac{t}{T_2}\right)^\nu} 
\end{equation}
where $\nu$ is the stretch factor set by the noise properties. The time evolution under Ramsey propagation is simulated with a generalized Lindblad master equation [\onlinecite{SM},\onlinecite{smirne2016ultimate}]. The sensing performance of optimized states is characterized by the square of the signal-noise-ratio SNR$^2 \propto \text{CFI}_\omega / t_\text{R}$ [\onlinecite{SM}]. Figure~\ref{fig:4}(d) shows their performance compared to the CSS and the GHZ states for a $\nu=2$ and $\nu=4$ noise exponent [\onlinecite{childress2006coherent}]. The created entangled states provide an advantage over uncorrelated states. For small spin numbers, the SNR follows the HL scaling [\onlinecite{chin2012quantum}].

\paragraph{Proposed experimental platforms.}

Candidate systems for realizing the proposed variational approach need to possess long $T_2$ coherence time, strong dipolar-interacting strength, and high initialization and readout fidelity. Recent developments in solid-state spin systems and ultracold molecules have demonstrated coherence times that exceed dipolar coupling times ($1/\bar{f}_{\text{dd}}$) as well as high-fidelity spin initialization and readout. Table~\ref{tab:table1} lists the experimentally observed parameters for different candidate systems, including Nitrogen Vacancy (NV) ensembles, P1 centers in diamond, rare-earth doped crystals, and ultracold molecule tweezer systems [\onlinecite{SM}].

\paragraph{Conclusion and Outlook.}

This work introduces a variational circuit that generates entangled metrological states in a dipolar-interacting spin system without requiring knowledge of the actual spin configuration. The required system parameters are within the reach of several experimental platforms. While this study remains limited to small system sizes ($N \leq 10$, limited by computational resource), our results are of immediate interest to nanoscale quantum sensing where spatial resolution is paramount and the finite sensor size limits the number of spins that can be utilized. Extending our results to $N>10$ can either be achieved by utilizing symmetries in regular arrays or directly testing our optimization algorithms on an actual experimental platform. The developed method is also potentially applicable for preparing other relevant highly entangled states in quantum computing and quantum communication.

We thank D. DeMille, D. Freedmann, A. Bleszynski Jayich, S. Kolkowitz, T. Li, Z. Li, R. Kaubruegger, Y. Huang, Q. Xuan, Z. Zhang, S. von Kugelgen, C-J. Yu, Y. Bao, and P. Gokhale for helpful discussions. T-X.Z., M.K., F.T.C., A.C., L.J. and P.M. acknowledge support by Q-NEXT Grant No. DOE 1F-60579. T-X.Z. and P.M. acknowledge support by National Science Foundation (NSF) Grant No. OMA-1936118 and OIA-2040520, and NSF QuBBE QLCI (NSF OMA- 2121044). S.Z. acknowledges funding provided by the Institute for Quantum Information and Matter, an NSF Physics Frontiers Center (NSF Grant PHY-1733907). Z.M. and F.T.C acknowledge support by EPiQC, an NSF Expedition in Computing, under grants CCF-1730082/1730449; in part by STAQ under grant NSF Phy-1818914; in part by the US DOE Office of Advanced Scientific Computing Research, Accelerated Research for Quantum Computing Program.; and in part by NSF OMA-2016136. The authors are also grateful for the support of the University of Chicago Research Computing Center for assistance with the numerical simulations carried out in this work.

Disclosure:  F.T.C. is the Chief Scientist for Super.tech and an advisor to QCI.

\let\oldaddcontentsline\addcontentsline
\renewcommand{\addcontentsline}[3]{}
\bibliography{main}
\let\addcontentsline\oldaddcontentsline



\newpage 
\begin{appendix}
\clearpage
\thispagestyle{empty}
\onecolumngrid
\begin{center}
\textbf{\large Supplemental Material for: Preparation of Metrological States in Dipolar Interacting Spin Systems}
\end{center}

\setcounter{equation}{0}
\setcounter{figure}{0}
\setcounter{table}{0}
\setcounter{page}{1}
\makeatletter
\renewcommand{\thefigure}{S\arabic{figure}}
\renewcommand{\thetable}{S\arabic{table}}
\renewcommand{\theequation}{S\arabic{equation}}

\tableofcontents

\newpage
\section*{Designing the variational circuit}
In this section, we discuss how to choose the experimentally realizable elementary gates in the variational sequence in the entangler based on limited quantum resource [\onlinecite{kaubruegger2019variational}, \onlinecite{kaubruegger2021quantum}].
\subsection*{Entanglement generation gates from two-body interaction Hamiltonian and global rotations}

Consider a two-body interaction Hamiltonian:

\begin{align} \label{eq:general_Hamiltonian}
     H_\text{int} = \sum_{i<j}V_{ij}\left(J^\text{I} S_{zi} S_{zj} + J^\text{S}\bm{S}_i \cdot \bm{S}_j\right).\numberthis
\end{align}
In this Hamiltonian, $\bm{S} = (S_x,S_y,S_z)$ is the vector of spin-1/2 operators, $V_{ij}$ is the interaction strength between spin $i$ and $j$ which depends on their locations, and $J^\text{I}(\neq0),J^\text{S}$ are the Ising and  symmetric coupling constant respectively.

The elementary gates in each layer of the variational circuit for preparing metrological states (Fig.1(c) main text) include two free evolutions under the interaction Hamiltonian $D(\tau), D(\tau')$, one global rotation along the $x$-axis $R_x(\vartheta)$ and two fixed $\pi/2$ rotations $R_y(-\frac{\pi}{2}),R_y(\frac{\pi}{2})$ along the $y$-axis. We define the interaction gate in the $z$-direction as

\begin{align} \label{eq:Dz}
     D_z(\tau) \equiv \exp(-i\tau H_\text{int}/\hbar) = \exp[-i\tau \sum_{i<j}V_{ij}\left(J^\text{I} S_{zi} S_{zj} + J^\text{S}\bm{S}_i \cdot \bm{S}_j\right)/\hbar].\numberthis
\end{align}

The interaction gates in other directions can be obtained by $\pi/2$ rotations:

\begin{align*} \label{eq:Dxy}
     D_{x,y}(\tau) &= R_{y,x}(\pi/2)D_z(\tau)R_{y,x}(-\pi/2)\\
     & = \exp[-i\tau \sum_{i<j}V_{ij}\left(J^\text{I} S_{x,yi} S_{x,yj} + J^\text{S}\bm{S}_i \cdot \bm{S}_j\right)/\hbar].\numberthis
\end{align*}
In Eqs.~(\ref{eq:Dz})~(\ref{eq:Dxy}), the symmetric interaction term stay unchanged because inner product is conserved under global rotation and the `direction of interaction' is only determined by the Ising term. Using these definitions, we simplify the gate set in each layer as

\begin{align*} \label{eq:DxRxDz}
    \mathcal{U}_i &= R_y\left(\frac{\pi}{2}\right)D\left(\tau'_i\right)R_y\left(-\frac{\pi}{2}\right)R_x\left(\vartheta_i\right)D\left(\tau_i\right)\\
    &= D_x\left(\tau'_i\right)R_x\left(\vartheta_i\right)D_z\left(\tau_i\right).\numberthis
\end{align*}
In the next two subsections, it will be shown that the sequence in Eq.~(\ref{eq:DxRxDz}) is the most general gate set that uses only global rotations and preserves the collective spin direction along $x$-direction.

\subsection*{Preservation of the collective spin direction}
Define the x-parity operator $P_x \equiv \Pi_i^N \sigma_{xi} = P_x^{\dagger}$, with $P_x^2 = I$. This operator describes the parity of a state in $x$-direction and is related to the global $\pi$ rotation along $x$-axis up to a phase constant, $R_x(\pi) = \exp(-i\pi\sum_iS_{xi}) = (-i)^N\Pi_i^N \sigma_{xi}$. Applying the x-parity operator onto individual spin's angular momentum operator gives $P_x S_{\mu j} P_x = (\sigma_x S_{\mu} \sigma_x)_j = \pm S_{\mu j}$. Thus the interaction gates along $x$- and $z$-direction conserve the x-parity, $P_x D_{x,z} P_x = D_{x,z}$. Similarly, the only global rotation  that conserves x-parity for arbitrary angles is $R_x(\vartheta)$. Then, based on Eq.(1) in the main text, the unitary operator of the whole control sequence conserves the x-parity

\begin{align*} \label{eq:whole_sequence}
    P_x\mathcal{S}(\bm{\theta})P_x &= P_x\mathcal{U}_m ... \mathcal{U}_2\mathcal{U}_1 P_x\\
    &= P_x \Pi_i [D_x\left(\tau'_i\right)R_x\left(\vartheta_i\right)D_z\left(\tau_i\right)]P_x\\
    & = \mathcal{S}(\bm{\theta}).\numberthis
\end{align*}

The initial spin state pointing to the $+x$-direction is an eigenstate of $P_x$: $P_x\ket{\uparrow_x}^{\otimes N} = \ket{\uparrow_x}^{\otimes N}$. Thus, any state produced by this variational circuit remains an eigenstate of $P_x$:

\begin{align*} \label{eq:Px_eigen}
    P_x\ket{\Psi(\bm{\theta})} &= P_x\mathcal{S}(\bm{\theta})\ket{\uparrow_x}^{\otimes N}\\
    &= P_x\mathcal{S}(\bm{\theta})P_x P_x\ket{\uparrow_x}^{\otimes N}\\
    &= \ket{\Psi(\bm{\theta})}.\numberthis
\end{align*}

Now consider the expectation value of the total spin angular momentum operator $J_{\mu} \equiv \sum_i S_{\mu i}$ $(\mu \in \{x,y,z\})$:
\begin{align*} \label{eq:JyJz}
    \langle J_{y,z}\rangle &= \bra{\Psi(\bm{\theta})}J_{y,z}\ket{\Psi(\bm{\theta})}\\
    &= \bra{\Psi(\bm{\theta})}P_xP_xJ_{y,z}P_xP_x\ket{\Psi(\bm{\theta})}\\
    &= -\langle J_{y,z}\rangle = 0.\numberthis
\end{align*}

Thus, the collective spin direction $\langle\bm{J}\rangle/|\langle\bm{J}\rangle|$ always points along the $x$-direction.

\subsection*{Choosing the most general gate set}
To preserve the collective spin direction along $x$-axis, the global rotation and interaction gates that can be chosen are $R_x$, $D_x$, $D_{\perp}$ where $D_{\perp}$ stands for the interaction gates along any direction perpendicular to the $x$-direction. Combining $R_x$ and $D_z$ can generate any $D_{\perp}$, thus the simplest gate set fulfilling all the requirements is $D_x R_x D_z$, as described by Eq.(1) in the main text.

The derivations and results in this section about selecting the variational sequence agree with ref.[\onlinecite{kaubruegger2019variational}]. However, the interaction Hamiltonian we discuss here is more general. In Eq.~(\ref{eq:general_Hamiltonian}), when $J^\text{I} = 1, J^\text{S} = 0$, the interaction becomes Ising type interaction which is equivalent to the Rydberg interaction in ref.[\onlinecite{kaubruegger2019variational},\onlinecite{kaubruegger2021quantum}]. The Ising interaction can also describe spin systems with large local disorder. The optimization results are shown in the next section. When $J^\text{I} = 3, J^\text{S} = -1$, Eq.~(\ref{eq:general_Hamiltonian}) becomes the dipolar interaction Hamiltonian between spin-1/2 particles. When $J^\text{I} = 2, J^\text{S} = -1$, it becomes the dipolar interaction Hamiltonian between spin-1 particles (such as NV centers). The simulation results for this case are shown in the next section. When $J^\text{I} = 1, J^\text{S} = -1$, the interaction can describe the dipolar interaction between cold molecules [\onlinecite{yan2013observation}]. 

\section*{Optimization algorithm: Covariance Matrix Adaptation Evolution Strategy}

The optimization in the $3m$ dimensional parameter space is highly non-convex (Fig.1(d) in main text) due to the large inhomogeneity of the interaction strength.
In our setting, the previously used Dividing Rectangles algorithm [\onlinecite{kaubruegger2019variational},\onlinecite{kokail2019self}] cannot converge to a beyond-SQL result despite large number of iterations.
We address this challenge by using the Covariance Matrix Adaptation Evolution Strategy (CMA-ES) as our optimization algorithm [\onlinecite{hansen2016cma}]. CMA-ES balances the exploration and exploitation process when searching in the parameter space so that convergence is reached after less than approximately 2,000 generations for $N,m \leq 10$. This corresponds to about $10^8$ repetitions of the Ramsey experiment, which can be further reduced if collective measurement observables are measured. 

We reduce the complexity of the optimization by restricting $\tau_i$ within $[0, 1/\bar{f}_\text{dd}]$ where $\bar{f}_\text{dd}$ is the average nearest-neighbor interaction strength for the considered spin configuration. Setting a large parameter searching range for the interaction gates' time $\tau_i$ would potentially ensure the global maximum CFI location is included in the parameter space. However, when the upper bound of $\tau_i$ is much bigger than $1/\bar{f}_\text{dd}$, the evolution of neighboring spin pairs is fast when sweeping $\tau_i$. This would introduce a huge amount of local maximum points in the parameter searching so that it is impractical for the black-box optimization algorithm to converge to that global maximum point.

\section*{Optimization results of different types of dipole-dipole interaction Hamiltonian}

 The magnetic dipole-dipole interaction Hamiltonian under secular approximation has the general form [\onlinecite{kucsko2018critical}, \onlinecite{ddinteraction}]:
\begin{align} \label{eq:HDD}
     H_\text{dd} = \sum_{i<j} V_{ij}  (2 S_{zi} S_{zj} - S_{xi} S_{xj} - S_{yi} S_{yj})\numberthis
\end{align}

with 
\begin{align} \label{eq:Vij}
     V_{ij} = \sum_{i<j}\frac{\mu_0}{4\pi}\frac{\gamma_i\gamma_j \hbar^2}{\left|\bm{r}_i - \bm{r}_j\right|^3}\frac{(1 - 3\cos\beta_{ij})}{2}\numberthis
\end{align}
where $\mu_0$ is the vacuum permeability, $\gamma$ is the geomagnetic ratio of the spin, $\beta_{ij}$ is the angle between the line segment connecting  ($\bm{r_i}$,$\bm{r_j}$) and the direction of the bias external magnetic field (along z-direction in this case). Eq.~(\ref{eq:HDD}) is able to describe the dipolar interaction for the spin systems with arbitrary spin number as long as the spin angular momentum operators $S_{\mu}$ obey the commutation relation $[S_{i},S_{j}] = i\epsilon_{ijk}S_{k}$. It applies to the spin-1/2 systems we discussed in the main text and Nitrogen-Vacancy (NV) centers which are spin-1 systems. 

\subsection*{NV ensemble}
Here we consider NV ensemble and only $\ket{m_s = 1}$ and $\ket{m_s = 0}$ are used as a 2-level system. The spin-1 operators are
\begin{align*}  \label{eq:S1}
    S_x^{(1)}=\frac{1}{\sqrt{2}}
    \begin{pmatrix}
    0 & 1 & 0\\
    1 & 0 & 1\\
    0 & 1 & 0\\
    \end{pmatrix},
    S_y^{(1)}=\frac{1}{\sqrt{2}}
    \begin{pmatrix}
    0 & -i & 0\\
    i & 0 & -i\\
    0 & i & 0\\
    \end{pmatrix}, 
    S_z^{(1)}= \begin{pmatrix}
    1 & 0 & 0\\
    0 & 0 & 0\\
    0 & 0 & -1\\
    \end{pmatrix}. \numberthis
\end{align*}

If we only take the $\ket{m_s = 1}$, $\ket{m_s = 0}$ subspace into consideration, the relations between the `truncated' spin-1 operators and the spin-1/2 operators are:

\begin{align}  \label{eq:S1 to S1/2}
    S_y^{(1)}=\sqrt{2}S_y^{(\frac{1}{2})},
    S_y^{(1)}=\sqrt{2}S_y^{(\frac{1}{2})},
    S_z^{(1)}= \frac{I}{2} + S_x^{(\frac{1}{2})} \numberthis
\end{align}

Plugging Eq.~(\ref{eq:S1 to S1/2}) into Eq.~(\ref{eq:HDD}), we get the effective dipole-dipole interaction Hamiltonian for NV ensemble $\ket{m_s = 1}$, $\ket{m_s = 0}$ subspace [\onlinecite{kucsko2018critical},\onlinecite{choi2017observation}]:

\begin{align}  \label{eq:HDD_NV}
   H_\text{DD,NV} = \sum_{i<j} V_{ij}  (S^{(\frac{1}{2})}_{zi} S^{(\frac{1}{2})}_{zj} - S^{(\frac{1}{2})}_{xi}S^{(\frac{1}{2})}_{xj} - S^{(\frac{1}{2})}_{yi}S^{(\frac{1}{2})}_{yj})\numberthis
\end{align}

\begin{figure}[h]
\includegraphics[width=150mm]{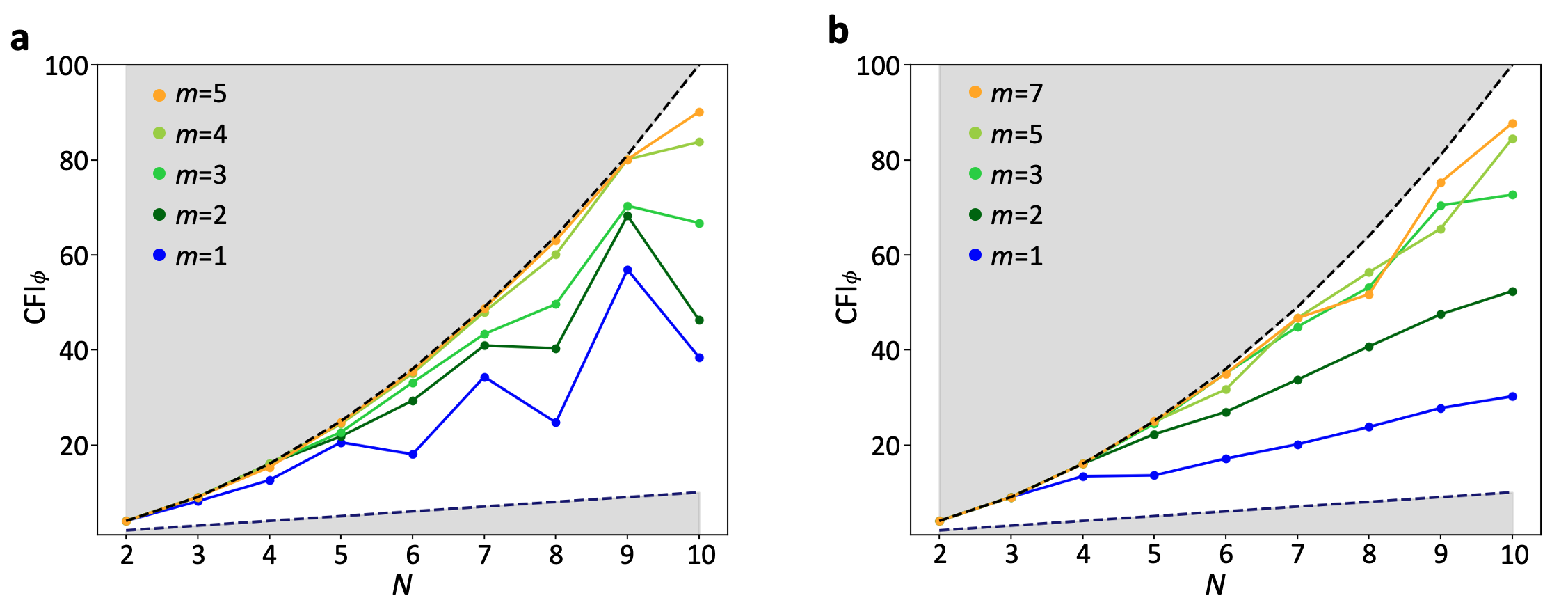}
\caption{\label{SI_fig:1} (a) Optimization results for NV-ensemble. The CFI saturates the theoretical upper bound Heisenberg Limit (HL) when the variational circuit layer number goes up from 1 to 7, and the CFI results are `oscillating' for even/odd number of spins from shallow circuit. (b) Optimization results for Ising type spin interaction when there is large local disorder in the system.}
\end{figure}

Fig.~\ref{SI_fig:1}(a) shows the Classical Fisher Information (CFI) optimization results for 2D square lattice spin configuration. They are similar to the results we get in Fig.(2) of the main text for spin-1/2 systems. 

\subsection*{Ising type interaction (large local disorder)}
When the system has large local disorder, the flip-flop terms in the dipolar interaction Hamiltonian Eq.~(\ref{eq:HDD}) are suppressed because of the large energy gap:

\begin{align*}  \label{eq:HDD_Ising}
   H_\text{DD,Ising} &= \sum_{i} \delta_i S^{(\frac{1}{2})}_{zi} + \sum_{i<j} V_{ij}  (2 S^{(\frac{1}{2})}_{zi} S^{(\frac{1}{2})}_{zj} - S^{(\frac{1}{2})}_{xi}S^{(\frac{1}{2})}_{xj} - S^{(\frac{1}{2})}_{yi}S^{(\frac{1}{2})}_{yj})\\
    &= \sum_{i} \delta_i S^{(\frac{1}{2})}_{zi} + \sum_{i<j} 2V_{ij}  ( S^{(\frac{1}{2})}_{zi} S^{(\frac{1}{2})}_{zj} - S^{(\frac{1}{2})}_{+i}S^{(\frac{1}{2})}_{-j} - S^{(\frac{1}{2})}_{-i}S^{(\frac{1}{2})}_{+j}) \\ 
    &\approx \sum_{i} \delta_i S^{(\frac{1}{2})}_{zi} + \sum_{i<j} 2V_{ij} S^{(\frac{1}{2})}_{zi} S^{(\frac{1}{2})}_{zj}. \numberthis
\end{align*}

This location-dependent single-spin energy shift ($\delta_i$) can be canceled by spin-echo pulse sequence where the interaction gate $D(\tau)$ needs to be applied:

\begin{align*}  \label{eq:spin_echo}
   D(\tau) 
   &= R_x(\pi)\exp[-i\tau H_\text{DD,Ising}]R_x(\pi)\exp[-i\tau H_\text{DD,Ising}]\\
   & = \exp[-i\tau \sum_{i<j} 2V_{ij} S^{(\frac{1}{2})}_{zi} S^{(\frac{1}{2})}_{zj}].\numberthis
\end{align*}

Eq.~(\ref{eq:spin_echo}) is also valid when the local disorder $\delta_i$ is comparable with the interaction strength $V_{ij}$. If there is local disorder in the dipolar-interacting spin ensemble, applying spin-echo will generate the interaction gate $D(\tau)$ where the local disorder terms are canceled.

The CFI optimization results by using the effective Ising type interaction Hamiltonian $H_\text{DD,Ising} = \sum_{i<j} 2V_{ij} S^{(\frac{1}{2})}_{zi} S^{(\frac{1}{2})}_{zj}$ is shown in Fig.~\ref{SI_fig:1} (b).

From Fig.~\ref{SI_fig:1}, the CFI results close to the Heisenberg Limit are observed, indicating that the variational method can be applied to different kinds of spins in solid state systems and generate highly entangled state for high-precision quantum metrology. We also observe that for shallow variational circuits, the CFI `oscillation' between even and odd spin numbers only appears when there are flip-flop terms in the Hamiltonian. For Ising type interaction, the `oscillation' disappears. 

\section*{Optimization results by using $P_z^{\text{tot}}$, $P_z^{\pi}$ as measurement bases}

\begin{figure}[h]
\includegraphics[width=150mm]{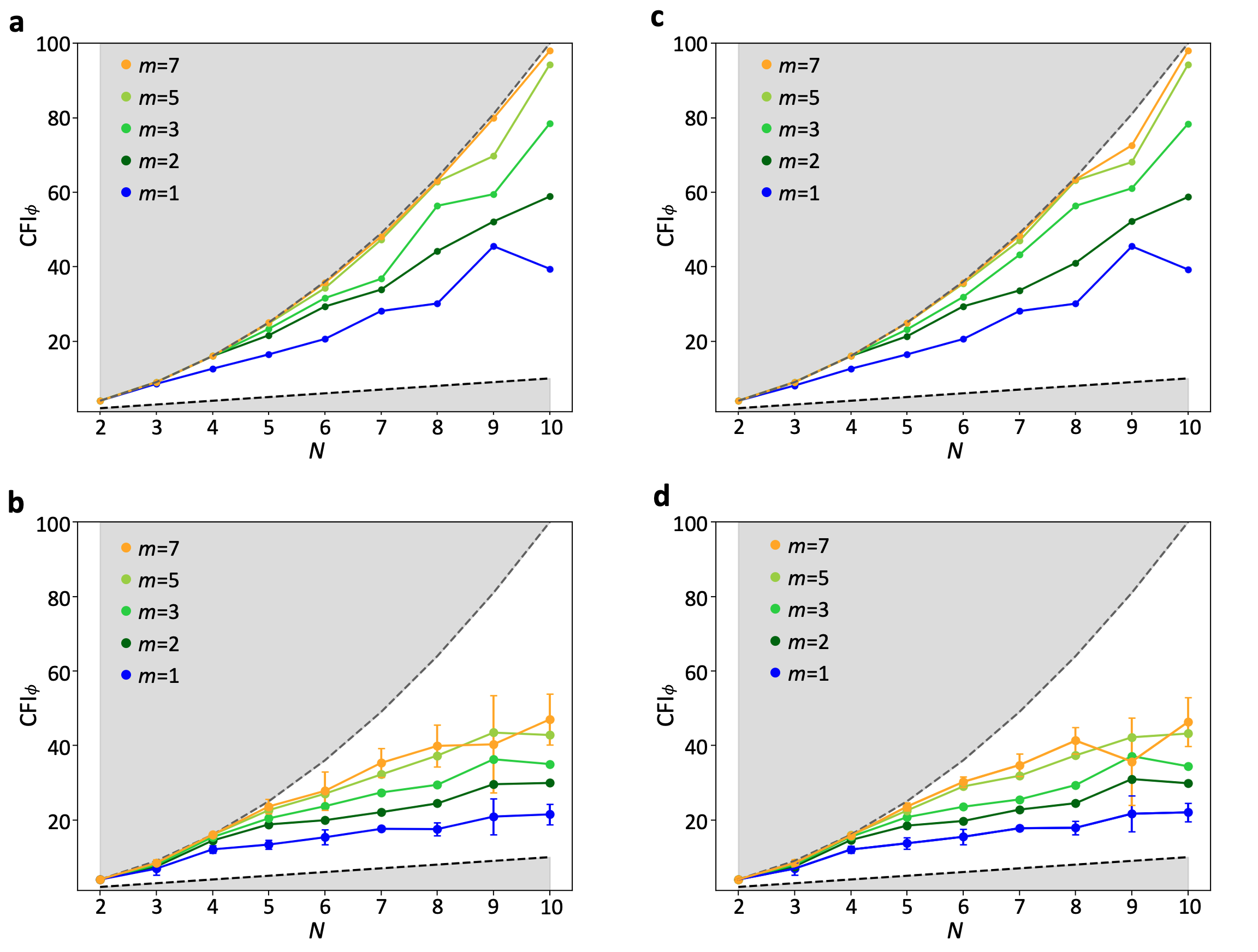}
\caption{CFI optimization data for (a) 2D square lattice using observable $P_z^{\text{tot}}$, (b) 3D-random configuration using observable $P_z^{\text{tot}}$ (averaged over 5 cases), (c) 2D square lattice using observable $P_z^{\pi}$, (d) 3D-random configuration using observable $P_z^{\pi}$ (averaged over 5 cases).}
\label{fig:cfi_Jz_parity}
\end{figure}

The optimization results shown in Fig.2 in the main text are obtain by using $P_z$ as the measurement basis for the CFI (cost function) calculation. Although measuring all the diagonal elements in the density matrix of the resulting states provides the maximum information one can get from single-qubit measurement and a large Hilbert space for the optimizer, it leads to an exponentially large ($2^N$) experimental repetition number when the CFI needs to be estimated from experimental data. Thus, we test the variational method on two other measurement bases which require less repetitions for readout.

The measurement basis on total spin polarization along $z$-direction is given by
\begin{align}\label{eq:PzTot}
    P_z^{\text{tot}} \equiv \ket{J=N/2,J_z}\bra{J=N/2,J_z}
    \numberthis
\end{align}
where $J$ is the total spin angular momentum quantum number and $J_z$ is the total spin angular momentum projection quantum number that runs from $N/2$ to $-N/2$. $P_z^{\text{tot}}$ has $N+1$ outcomes, so it scales linear with the system size. 

The optimization results by using the CFI on $P_z^{\text{tot}}$ as cost function are shown in Fig.~\ref{fig:cfi_Jz_parity} (a)(b). Surprisingly, compared to the results by using $P_z$, the optimization results from using $P_z^{\text{tot}}$ are improved by about a factor of $1.5\sim2$ for the 3D-random spin configuration. Since all the information one can extract from $P_z^{\text{tot}}$ are contained in $P_z$, we attribute this improvement to the simpler parameter space structure that $P_z^{\text{tot}}$ provides to the optimizer. Less local maximum points in the parameter space will help the optimizer to converge to a high CFI point, especially when the dimension of the parameter space ($3m$) is large.

Parity of the spin ensemble, 
\begin{align}\label{eq:Parity}
    P_z^{\pi} \equiv \Pi_i^N \sigma_{zi},
    \numberthis
\end{align}
provides a constant ($2$) dimensional outcome space for experimental readout. Improvements are also observed in 2D square lattice and 3D-random spin configurations (Fig.~\ref{fig:cfi_Jz_parity}(c)(d)).

\section*{Supplementary Data}

\subsection*{Complete CFI data for Fig.2 in main text}

\begin{figure}[h]
\includegraphics[width=150mm]{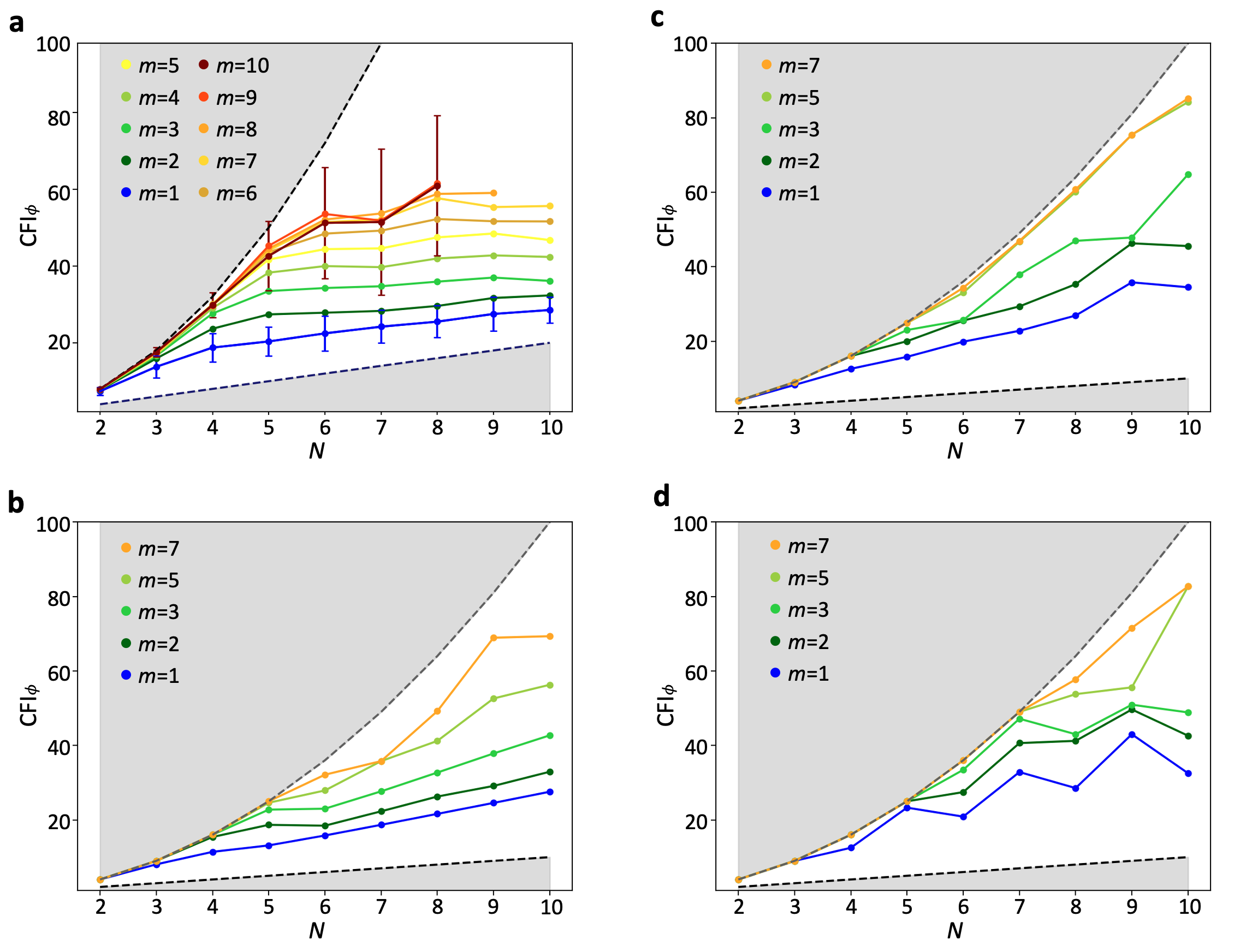}
\caption{ The complete CFI data for (a) 3D random, (b) 1D chain, (c) 2D square lattice, and (d) 2D circle.}
\label{fig:complete_CFI_data}
\end{figure}

The complete data for dipolar-interacting spin systems' CFI optimization is shown in this section. Fig.~\ref{fig:complete_CFI_data}(a) shows the 50-cases averaged optimization results for 3D-random spin configurations, the variational circuit layer number $m$ is chosen from 1 to 10. The optimized CFI results are approaching to the Heisenberg Limit (HL) when more layers ($m$) are used. However, when $m>7$, the CFI results stop increasing. This CFI `saturation' effect might be caused by two reasons. First, when $m$ is larger, the number of the local maximum points in the high dimensional parameter space increases. This could potentially cause the optimizer to stuck in the local maximum point. Sometimes, take $N=7, m=10$ data in Fig.~\ref{fig:complete_CFI_data}(a) as an example, adding more variational layers even leads to a lower CFI optimization result. The `local maximum' problem could be solved by more advanced and powerful optimization algorithms, such as reinforcement learning [\onlinecite{sutton2018reinforcement,silver2017mastering,peng2021deep}], and more computational resources. Second, the `saturation' effect reflects the global maximum CFI one can reach, no matter what kind of optimization algorithm is applied. It's still an open question what is the highest CFI the spin ensemble could reach for a given configuration.

Fig.~\ref{fig:complete_CFI_data}(b)-(d) show the CFI optimization result for 1D chain, 2D square lattice and 2D symmetric cycle spin configurations. The results of regular patterns are better than those of 3D-random pattern.

\newpage

\subsection*{Required layers to reach given CFI for 2D square lattice}
\begin{figure}[h]
\includegraphics[width=75mm]{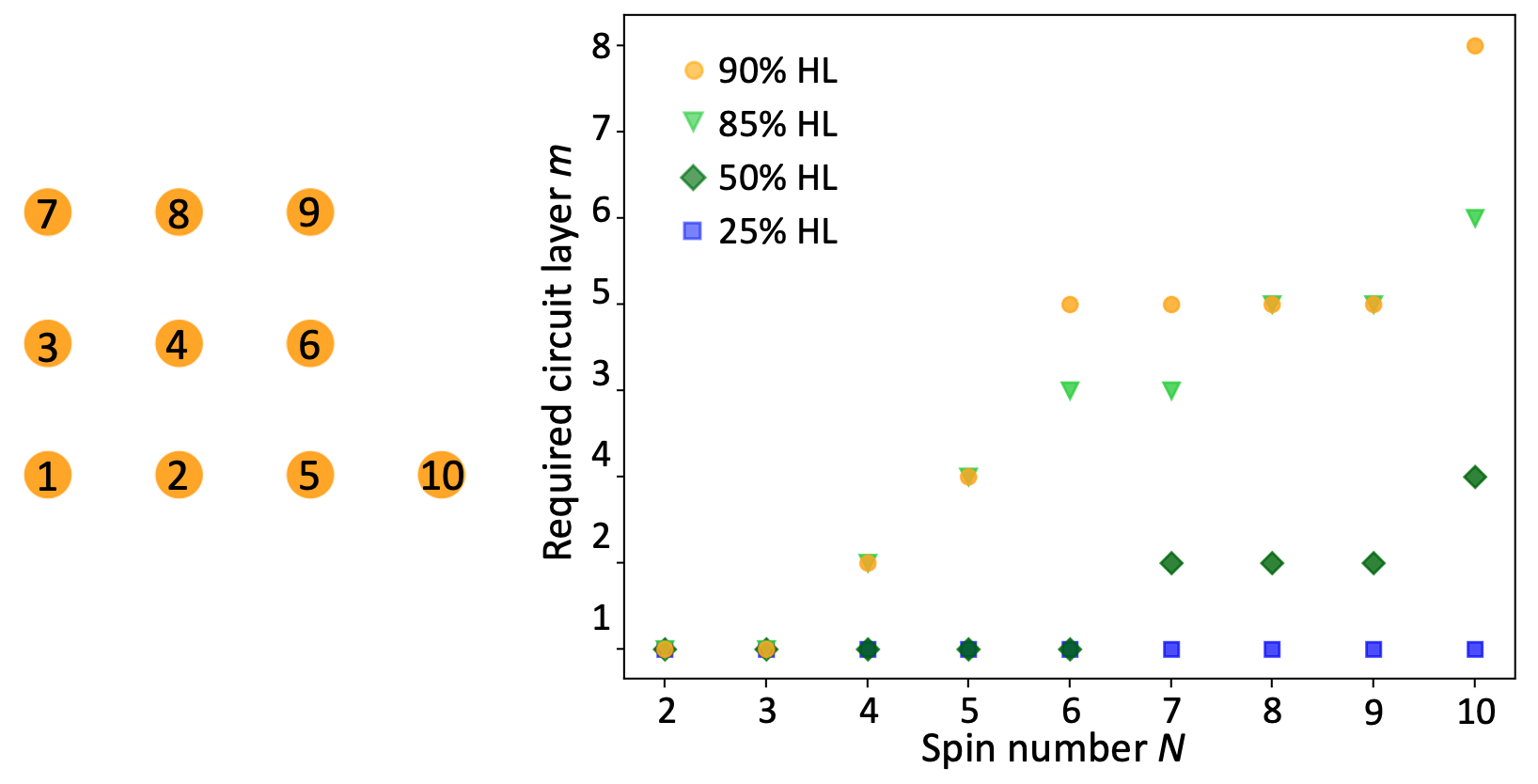}
\caption{Left: Schematic of a 2D square lattice pattern. The numbers label the order in which spins are added to form a lattice of size $N$. Right: Number of layers required to achieve a CFI within a given percentage of the HL.}
\label{fig_scalability}
\end{figure}
As shown by the schematic on the left, the distances between spin No.4 and spin No.5, 7, and 9 are the same, so the interaction strengths between each pair are the same. Similarly, the distance between spin No.4 and spin No.2, 3, 6, and 8 are the same (smaller). Therefore, the plateau features in Fig.~\ref{fig_scalability} are likely due to this symmetry: adding one more spin to the lattice does not require an extra layer to reach a given percentage of the CFI. 

\subsection*{Orders of interaction}
\begin{figure}[h]
\includegraphics[width=75mm]{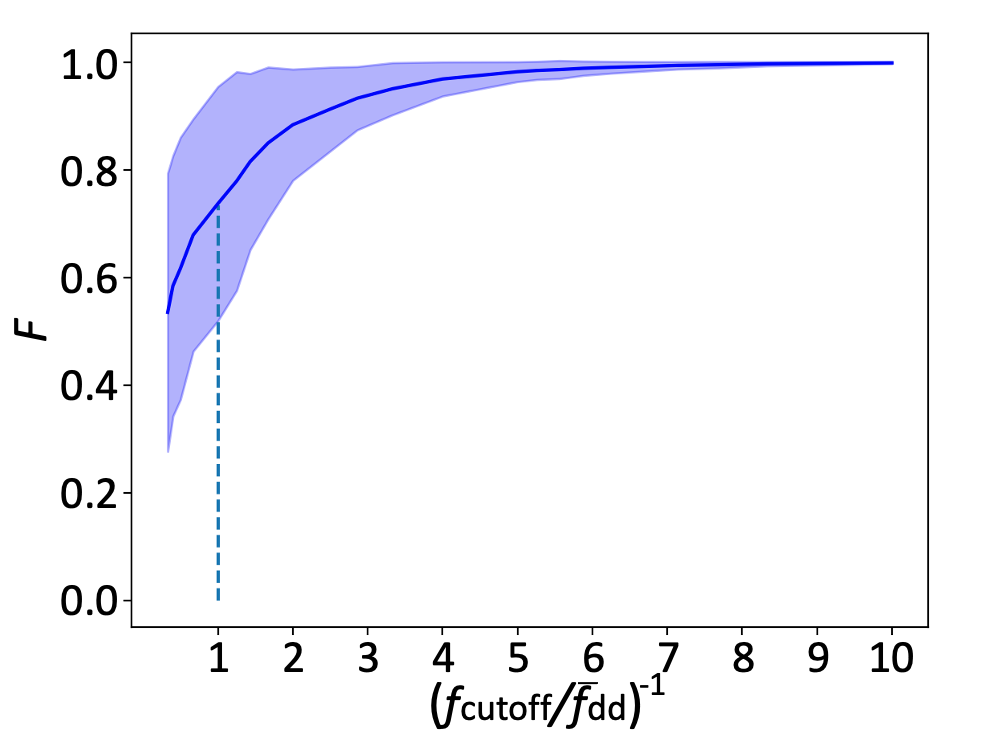}
\caption{Average state fidelity vs. different cutoff strength in $H_\text{dd}$. The shaded area corresponds to the error range. Data obtained from 3D-random $N = 10, m = 5$, 50-cases optimization results.}
\label{fig_interaction}
\end{figure}

Due to the decaying feature ($\frac{1}{r^3}$) of dipolar interaction strength, the resulting states might be mainly generated by nearest-neighbor interaction. For studying `how much' interaction is essential for generating the resulting entangled states, we calculate the overlap (state fidelity [\onlinecite{nielsen_chuang_2010}]) between the original state and the new state, which is generated by using the cutoff Hamiltonian and optimized parameters. A cutoff interaction strength $f_{\text{cutoff}}$ is chosen, and all the pairwise potential $V_{ij}$ smaller than $f_{\text{cutoff}}$ are set to zero in the cutoff Hamiltonian.  Fig.~\ref{fig_interaction} shows the relation between the state fidelity $F$ versus $f_{\text{cutoff}}$. A state fidelity value less than 1 is observed when $f_{\text{cutoff}}$ is set to be equal to the averaged nearest-neighbor interaction strength $f_{\text{dd}}$. This result reflects higher order interactions in the spins ensemble are utilized for the entangled state generation.

\subsection*{Non-Markovian noise sensing performance}
\begin{figure}[h]
\includegraphics[width=75mm]{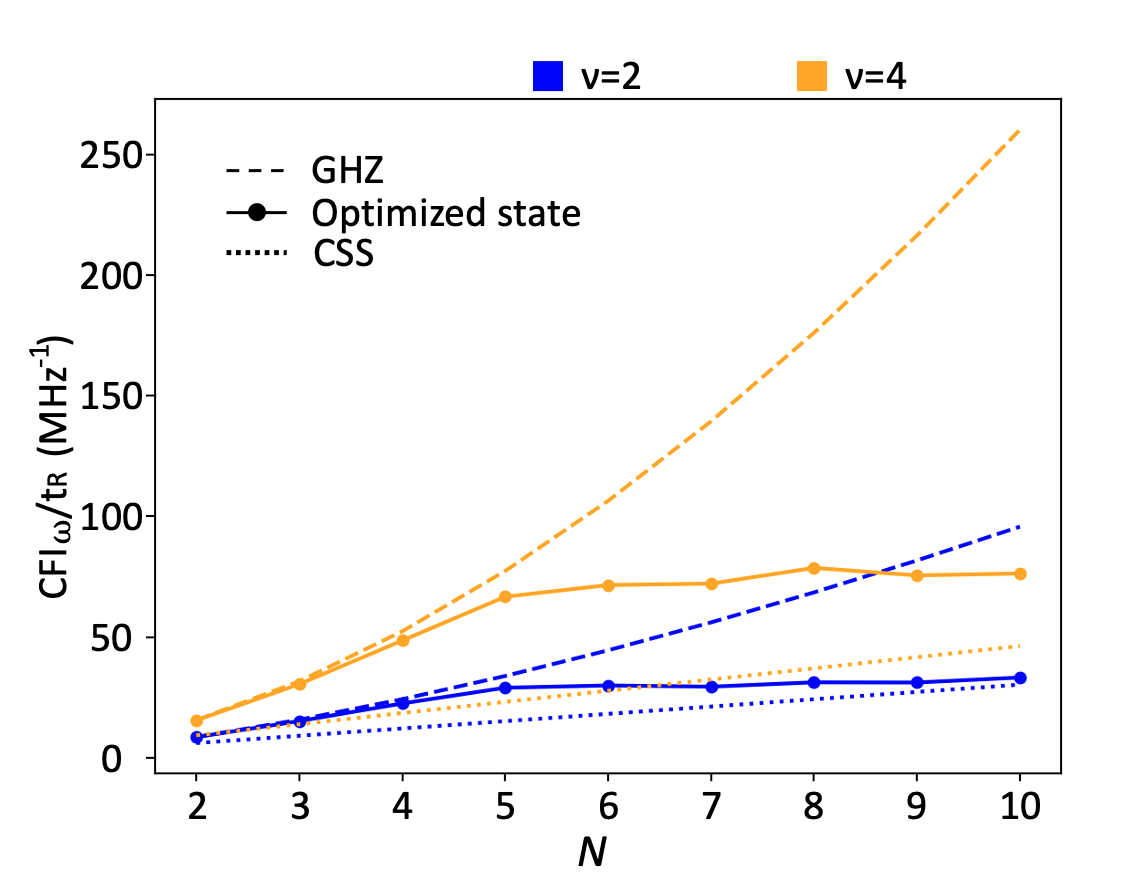}
\caption{Average Ramsey protocol's results of the generated entangled states in 3D random configurations when considering non-Markovian noise in the signal accumulation step. Blue and orange correspond to two different noise models ($\nu=2$ and $4$).}
\label{figS3}
\end{figure}

\subsection*{Optimized states with different readout fidelity}

We run the optimization with imperfect readout for $N=4$ and $N=10$ 2D square lattice spin configurations. The optimized states resemble GHZ states (high RF), SSS (low RF), CSS (RF close to 50\%). For $N=4$ case, the Gaussian state appears for RF lower than 92\%, but for $N=10$ case, the Gaussian states appears when RF is about 96\%. We expected that for large spin system with finite RF, Gaussian states (e.g. SSS) are advantageous for quantum-enhanced metrology.

\begin{figure}[h]
\includegraphics[width=150mm]{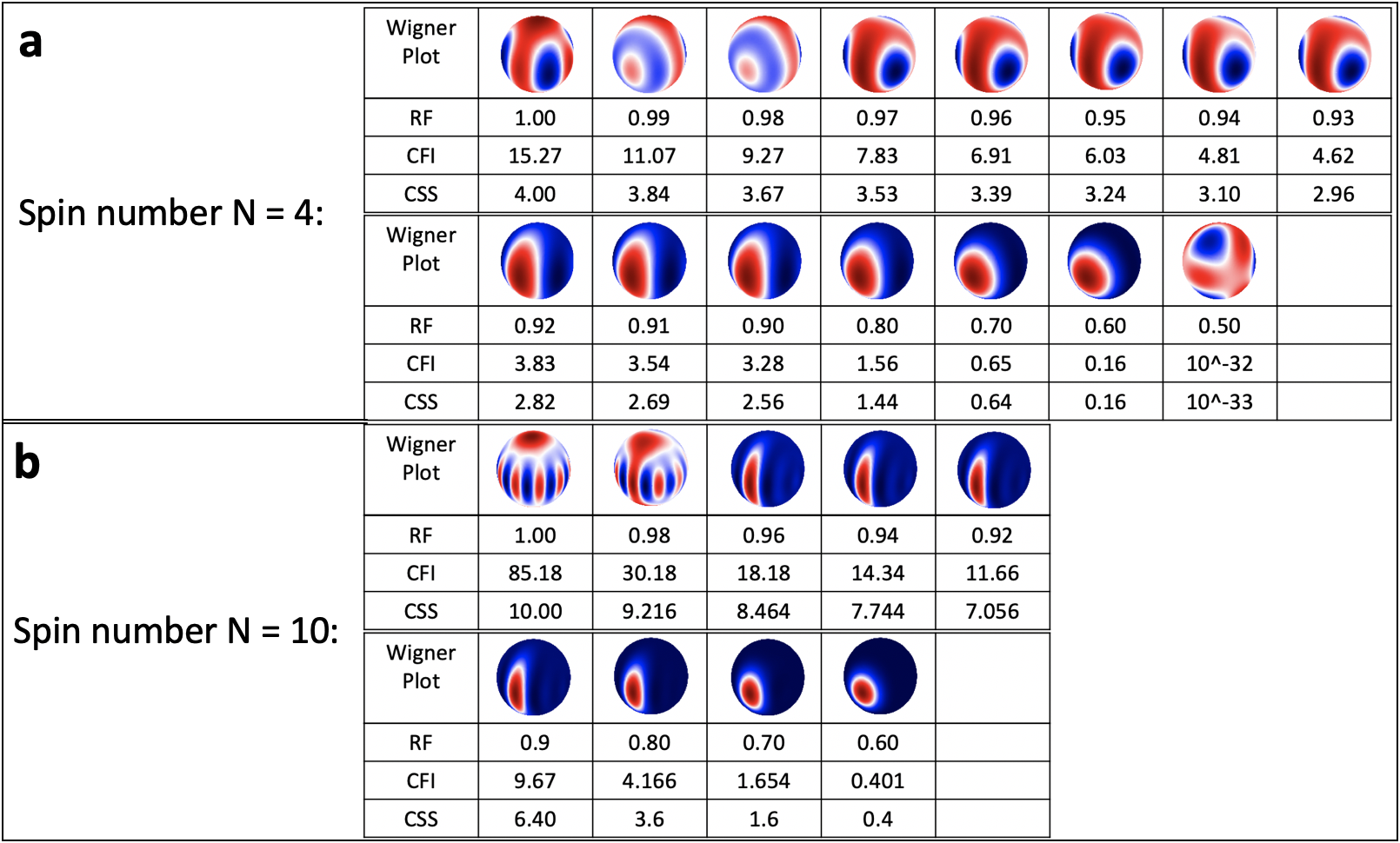}
\caption{Optimized states' Wigner distributions when finite RF is assumed in the CFI calculation.}
\label{fig:Optimization_with_RF}
\end{figure}

\subsection*{Relative experimental parameter table (full)}

\begin{table*}[h]
\caption{\label{tab:table1}%
Experimental platforms' data}
\begin{ruledtabular}
\begin{tabular}{c|lllllll}
System & $T_2^\text{(best)}$ & $T_2^\text{(DD)}$ & $\bar{f}_\text{dd}$ & $P_\text{ini}$ & $F_\text{readout}$ & $\nu$ \\
\hline
NV ensemble    & $1.58(7)$s\textsuperscript{a} &$7.9(2)\mu$s\textsuperscript{b}    & $35$kHz\textsuperscript{b}  & $97.5\%$\textsuperscript{c} & $97.5\%$\textsuperscript{c} & $2-4$\textsuperscript{b}  \\
P1 centers    & $0.8$ms\textsuperscript{e}(DEER) &  $4.4\mu$s\textsuperscript{f}  & $0.7$kHz\textsuperscript{e},$0.92$MHz\textsuperscript{f}  & $95\%$\textsuperscript{e} & $95\%$\textsuperscript{e} & ?  \\
Rare-Earth crystals & $23.2\pm0.5ms$\textsuperscript{g}  & $2.5\mu $s\textsuperscript{h}    & $1.96$MHz\textsuperscript{h}  & $97\%$\textsuperscript{i} & $94.6\%$\textsuperscript{j} & $2.4\pm0.1^{\text{g}}$  \\
Cold Molecules & $~1$s\textsuperscript{k}  & $80$ms\textsuperscript{l}    & $52$Hz\textsuperscript{l},$1.5$kHz\textsuperscript{m}  & $97\%^{\text{m}}$ & $97\%^{\text{m}}$ & ?  \\
\end{tabular}
\end{ruledtabular}
\footnotesize{$^\text{a}$ T.H.Taminiau, NComm 2018, $^\text{b}$ H.Zhou, PRX 2020, B.J.Shields, $^\text{c}$ M.D.Lukin, PRL 2015, $^\text{d}$ L.Childress Science 2006 \\}
\footnotesize{$^\text{e}$ T.H.Taminiau, NComm 2021, $^\text{f}$ N.Yao, Nature 2021\\}
\footnotesize{$^\text{e}$ P.Bertet, Science advances 2021, $^\text{h}$ A.Reiserer, PRL 2021, $^\text{i}$ J.Thompson, Science 2020, $^\text{j}$ J.Thompson, NComm 2020\\}
\footnotesize{$^\text{k}$ M.R Tarbutt PRL 2020, $^\text{l}$ B.Yan, J.Ye, Nature 2013, $^\text{m}$ J Doyle, PRL 2020}

\end{table*}

Based on the simulation results shown in Fig.4(c) in main text, we need $\bar{f}_{dd} T_2 \geq 5$ to generate metrological states that beat the SQL. It's worth mentioning that the $T_2$ in this situation stands for the coherence time \textit{without} the dipole-dipole interaction's influence. During the state preparation step, the dipolar interactions between the spins are included in the system Hamiltonian for the entanglement generation ($D$ gate in Fig.1(c) in main text). Thus, the $T_2$\textsuperscript{(DD)} in Table~\ref{tab:table1} is a lower bound and $T_2$\textsuperscript{(best)} is a more precise estimation for the spin coherence time.


\section*{Supplementary Derivations}

\subsection*{CFI with respect to angle and frequency}
In general, the Classical Fisher Information (CFI) measures the sensitivity of a statistical model to small changes of a parameter $\theta$ [\onlinecite{braunstein1994statistical}, \onlinecite{strobel2014fisher}]. Let $Z$ be a random variable and $P_z(\theta) \equiv P(z|\theta)$ be its probability distribution which depends on $\theta$. Let $\Theta$ be an unbiased estimator of $\theta$, i.e.
\begin{align} \label{eq:20}
    \theta = \langle \Theta \rangle=\sum_z \Theta\cdot P_z(\theta).\numberthis
\end{align}
From Eq.~(\ref{eq:20}) and the fact that the sum of probabilities of all outcomes is $1$, 
\begin{align}
    1 &= \frac{\partial \langle \Theta \rangle}{\partial \theta} = \frac{\partial}{\partial \theta}\sum_z \Theta P_z(\theta)\label{eq:21},\numberthis\\
    0 &= \frac{\partial}{\partial \theta}\sum_z P_z(\theta) \label{eq:22}.\numberthis 
\end{align}
Subtracting Eq.~(\ref{eq:22}) multiplied by $\theta$ from Eq.~(\ref{eq:21}), we get
\begin{align*}\label{eq:23}
    1&=\sum_z (\Theta-\theta)\frac{\partial}{\partial \theta} P_z(\theta)\\
    &=\sum_z P_z(\theta) (\Theta-\theta)\frac{1}{P_z(\theta)} \frac{\partial}{\partial \theta}P_z(\theta)\\
    &=\langle (\Theta-\theta) \frac{1}{P_z(\theta)} \frac{\partial}{\partial \theta}P_z(\theta)\rangle\\
    &=\numberthis\langle (\Theta-\theta) \frac{\partial}{\partial \theta} \log P_z(\theta)\rangle.
\end{align*}
Letting $X=\Theta-\theta$ and $Y=\frac{\partial}{\partial \theta} \log P_z(\theta)$, by the Cauchy-Schwartz inequality for random variables: $\langle XY \rangle^2\leq \langle X^2 \rangle \langle Y^2 \rangle$, we have 
\begin{align}\label{eq:24}
    \langle (\Theta-\theta)^2\rangle \bigg \langle \left( \frac{\partial}{\partial \theta} \log P_z(\theta)\right)^2 \bigg \rangle \geq 1,\numberthis
\end{align}
where 
\begin{align*} \label{eq:25}
    \langle (\Theta-\theta)^2\rangle &= \langle \Theta^2\rangle -(2\theta\langle \Theta\rangle -\langle\theta^2\rangle)\\
    &= \langle \Theta^2 \rangle -(2\theta^2-\theta^2)\\
    &= \langle \Theta^2 \rangle - \langle \Theta\rangle^2\\
    &= \numberthis\Delta \Theta^2
\end{align*} is the variance of $\Theta$. Defining 
\begin{align} \label{eq:26}
    \mathrm{CFI}=\sum_z P_z(\theta)\left(\frac{\partial}{\partial \theta} \log P_z(\theta)\right)^2,\numberthis
\end{align}
we have
\begin{align} \label{eq:27}
    \Delta \Theta^2 \geq \frac{1}{\mathrm{CFI}}.\numberthis
\end{align}
If the measurement is repeated $M$ times, then by the additive property of CFI, we obtain the Cram\'{e}r-Rao bound:
\begin{equation} \label{eq:28}
    \Delta \Theta^2 \geq \frac{1}{M\cdot\mathrm{CFI}}.\numberthis
\end{equation}

In our variational circuit, we use CFI with respect to an infinitesimal angle $\phi$ as the cost function to generate entangled states. In our program, we use a method similar to parameter shift to calculate the $\mathrm{CFI}_\phi$ of our optimized states [\onlinecite{cerezo2021variational},\onlinecite{meyer2021variational},\onlinecite{schuld2019evaluating}]. In the following notation,
\begin{enumerate}
    \item $z$ represents a multi-qubit state in the z-basis;
    \item $\mathcal{U}(\phi)=e^{-i\phi J_y}$ is the rotation operator where $\phi$ is a small angle;
    \item $\psi$ is the state we create from the variational circuit;
    \item $P_z(\theta)$ is the probability of measuring the state $z$ with the state after rotation.
\end{enumerate}
Then 
\begin{align*} \label{eq:29}
    \frac{\partial}{\partial \phi} P_z(\phi)\Bigr\rvert_{\phi \rightarrow 0}=& \frac{\partial}{\partial \phi} |\langle z|\mathcal{U}(\phi)|\psi\rangle|^2\Bigr\rvert_{\phi \rightarrow 0}\nonumber\\
    =& \frac{\partial}{\partial \phi} \langle \psi|\mathcal{U}^\dagger (\phi) |z\rangle \langle z|\mathcal{U}(\phi)|\psi\rangle\Bigr\rvert_{\phi \rightarrow 0}\nonumber\\
    =& \langle \psi|\mathcal{U}^\dagger (\phi) i J_y |z\rangle 
    \langle z|\mathcal{U}(\phi)|\psi\rangle\Bigr\rvert_{\phi \rightarrow 0}+ \langle \psi|\mathcal{U}^\dagger (\phi) |z\rangle 
    \langle z|\mathcal{U}(\phi)(-i)J_y|\psi\rangle\Bigr\rvert_{\phi \rightarrow 0}\nonumber\\
    =&i\bra{\psi}\bigg(J_y\ket{z}\bra{z}-\ket{z}\bra{z}J_y\bigg)\ket{\psi}.\numberthis
\end{align*}
Note that assuming the rotation operator $\mathcal{U}(\phi)=e^{-i\phi J_y} \equiv \mathcal{U}_y(\phi)$ along $y$-axis is for calculation simplicity. In experiments, the signal (e.g. the external B-field) usually induces a rotation along $z$-axis, $\mathcal{U}_z(\phi) = e^{-i\phi J_z}$. It's equivalent to assume that the prepared state is firstly rotated by a $R_x(\pi/2)$ pulse and then accumulates a signal $\phi$ along $y$-axis, or firstly accumulates a signal along $z$-axis and then rotated by $R_x(-\pi/2)$ pulse [\onlinecite{strobel2014fisher}]. In another word, $R_x(-\pi/2)\mathcal{U}_z(\phi) = \mathcal{U}_y(\phi) R_x(\pi/2)$, so the signal accumulation process we assumed in the calculation is able to simulate the experiments.

After creating the entangled states, we want to know how useful they are in a Ramsey spectroscopy, where the signal we want to detect is a frequency $\omega$. By the same calculation as above except the difference that we take derivative with respect to $\omega=\frac{\phi}{t_\text{R}}$ where $t_\text{R}$ is the Ramsey sensing time, we have 
\begin{align}  \label{eq:30}
    \mathrm{CFI}_\omega = \mathrm{CFI}_\phi \cdot t_\text{R}^2.\numberthis
\end{align}

\subsection*{Relation beteen $\text{CFI}_\omega$ and SNR in single qubit Ramsey experiment}
We illustrate the Ramsey protocol for a single qubit.
\begin{enumerate}
    \item The qubit is initialized into the ground state $\ket{0}$.
    \item A $\frac{\pi}{2}$ pulse along the y-direction is applied to transform it into a superposition state $\frac{1}{\sqrt{2}}(\ket{0}+\ket{1})$. Its matrix form is
\begin{align}  \label{eq:31}
    \rho(t)=\frac{1}{2}
    \begin{pmatrix}
    1 & 1\\
    1 & 1
    \end{pmatrix}.\numberthis
\end{align}
    \item After evolving under noise and a signal with frequency $\omega$ for time $t$, its state becomes
\begin{align} \label{eq:32}
\rho(t)=
    \frac{1}{2}
    \begin{pmatrix}
1 & e^{-i\omega t}e^{-2 \gamma t}\\
e^{i\omega t}e^{-2 \gamma t} & 1
\end{pmatrix}\numberthis
\end{align}
where $\gamma$ is the decoherence rate.
    \item A second $\frac{\pi}{2}$ pulse along the x-direction is applied for readout. The qubit is then in the state
\begin{align} \label{eq:33}
R_x\left(\frac{\pi}{2}\right)\rho(t)R_x^\dagger\left(\frac{\pi}{2}\right)\numberthis
\end{align}.
    \item After the rotation, the probability of the qubit being in the ground state is
\begin{align}  \label{eq:34}
P_0=
\frac{1}{2}+\frac{1}{2}e^{-2\gamma t}\sin{\omega t}.\numberthis
\end{align}
\end{enumerate}

The CFI with respect to $\omega$ is
\begin{align*} \label{eq:35}
    \mathrm{CFI}_\omega&=\frac{1}{P_0}\left(\frac{\partial P_0}{\partial \omega}\right)^2 +\frac{1}{P_1}\left(\frac{\partial P_1}{\partial \omega}\right)^2=\frac{t^2\cos^2{\omega t}}{e^{4\gamma t}-\sin^2{\omega t}}.\numberthis
\end{align*}
Assuming only quantum projection noise, the Signal-to-Noise Ratio (SNR) is $\frac{\delta P_0}{\sqrt{\frac{1}{M}P_0(1-P_0)}}$ where $M$ is the total number of measurements. Then 
\begin{align}  \label{eq:36}
    \mathrm{SNR}^2=\frac{M t^2 \cos^2{\omega t} \delta \omega^2}{e^{4\gamma t}-\sin^2{\omega t}}.\numberthis
\end{align}
Assuming no time overhead, i.e., $M=\frac{T_\text{tot}}{t_\text{R}}$ where $T_\text{tot}$ is the total measurement time and $t_\text{R}$ is the time between Ramsey pulses,
we obtain the relationship
\begin{equation} \label{eq:37}
\mathrm{CFI}_\omega \cdot \frac{T_\text{tot}}{t_\text{R}} \cdot\delta\omega^2=\mathrm{SNR}^2.\numberthis
\end{equation}
In unit time ($T_\text{tot}=1$), when SNR$=1$, the smallest signal we can measure is 
\begin{align} \label{eq:38}
    \delta\omega = \frac{1}{\sqrt{M\cdot \mathrm{CFI}_\omega}},\numberthis
\end{align}
leading to the saturated Cram\'{e}r-Rao bound.

\subsection*{Maximum Likelihood Estimator}
 Since a measurement collapses a quantum state to an eigenstate of the observable, it's impossible to directly measure $P(\theta)$. In experiments, we repeat the sequence to obtain the results for estimating the $P(\theta)$ and then get an estimate value of $\theta$. To understand the relation between the variance of the estimation and CFI, we introduce the Maximum Likelihood Estimator (MLE), which has asymptotic properties to saturate the Cram\'{e}r-Rao bound. Below we summarize the proof given in [\onlinecite{mle}].

Let $\bm{X}=\{X_1, X_2, ..., X_M\}$ be a collection of independent and identically distributed (i.i.d.) random variables with a parametric family of probability distributions $\{P(X|\theta)|\theta\in\Theta\}$, where $\theta$ is an unknown parameter and $\Theta$ is the parameter space. Let $\bm{x}=\{x_1,x_2,...,x_M|x_i\in X_i\}$ be the experimental data set from $M$ repetitions. The goal is to estimate $\theta$ (the signal we want to measure) from $\bm{x}$, i.e., find $\theta$ that is most likely to produce the outcome $\bm{x}$. Thus, the normalized log-likelihood function is defined as
\begin{align*} \label{eq:39}
L_M(\theta)&=\frac{1}{M}\log P(\bm{X}|\theta)=\frac{1}{M}\log \prod_{i=1}^M P(X_i|\theta)=\frac{1}{M}\sum_{i=1}^M \log P(X_i|\theta).\numberthis
\end{align*}
A MLE maximizes the log-likelihood function
\begin{equation} \label{eq:40}
\Theta_{\mathrm{MLE}} = \argmax_{\theta\in\Theta} L_M(\theta).\numberthis
\end{equation}
In the following, we first show that 
\begin{enumerate}
    \item $\Theta_{\mathrm{MLE}}$ converges to the true parameter $\theta_0$;
    \item the distribution of $\sqrt{M}(\Theta_{\mathrm{MLE}}-\theta_0)$ tends to a normal distribution $\mathcal{N}\left(0,\frac{1}{\mathrm{CFI}_{\theta_0}}\right)$ as $M$ increases. 
\end{enumerate}
In other words, not only does the MLE converge to the true parameter, it converges at a rate $\frac{1}{\sqrt M}$.

Define
\begin{align} \label{eq:41}
    L(\theta)=\langle \log P(\bm{X}|\theta)\rangle_{\theta_0}\numberthis
\end{align}
which denotes the expected log-likelihood function with respect to $\theta_0$, then by the Weak Law of Large Numbers (WLLN), the average outcomes from a large number of trials should approach the expected value:
\begin{equation} \label{eq:42}
\forall \theta, L_M(\theta) \xrightarrow{\mathit{M\rightarrow \infty}} L(\theta).\numberthis
\end{equation}
In fact, $\theta_0$ maximizes $L(\theta)$:
\begin{align*} \label{eq:43}
   \forall \theta, L(\theta)-L(\theta_0) &=\langle \log P(\bm{X}|\theta)\rangle_{\theta_0} - \langle \log P(\bm{X}|\theta_0)\rangle_{\theta_0}\nonumber\\
   &=\bigg\langle\log \frac{P(\bm{X}|\theta)}{P(\bm{X}|\theta_0)}\bigg\rangle_{\theta_0}\nonumber\\
   &\leq \bigg \langle\frac{P(\bm{X}|\theta)}{P(\bm{X}|\theta_0)}-1 \bigg \rangle_{\theta_0}\nonumber\\
   &=\sum_{\bm{x}\in\bm{X}} \left(\frac{P(\bm{x}|\theta)}{P(\bm{x}|\theta_0)}-1\right)P(\bm{x}|\theta_0)\\
   &=1-1=0.\numberthis
\end{align*}
Moreover, we show that $\theta_0$ is the unique maximizer. Jensen's inequality states that for a strictly convex function $f$ and a random variable $Y$, 
\begin{align} \label{eq:44}
    \langle f(Y) \rangle > f(\langle Y \rangle).\numberthis
\end{align}
Taking $f(y)=-\log y$ and $P(\bm{X}|\theta)\neq P(\bm{X}|\theta_0)$, we have
\begin{equation} \label{eq:45}
    \bigg\langle -\log \frac{P(\bm{X}|\theta)}{P(\bm{X}|\theta_0)}
\bigg\rangle_{\theta_0}>-\log \bigg \langle\frac{P(\bm{X}|\theta)}{P(\bm{X}|\theta_0)} \bigg \rangle_{\theta_0}=0,\numberthis
\end{equation}
or 
\begin{equation} \label{eq:46}
    L(\theta_0)>L(\theta).\numberthis
\end{equation}
Therefore, since 
\begin{enumerate} 
    \item $\Theta_{\mathrm{MLE}}$ maximizes $L_M(\theta)$,
    \item $\theta_0$ maximizes $L(\theta)$, and
    \item $L_M(\theta) \xrightarrow{\mathit{M\rightarrow \infty}} L(\theta)$,
\end{enumerate}
$\Theta_{\mathrm{MLE}}$ converges to $\theta_0$.

Now we use this property to prove that the distribution of $\Theta_{\mathrm{MLE}}$ tends to the desired normal distribution, where we will apply the Central Limit Theorem (CLT):
Suppose $\bm{X}=\{X_1,...,X_M\}$ is a sequence of i.i.d. random variables with $\langle X_i\rangle=\mu$ and $\mathrm{Var}(X_i)=\sigma^2<\infty$. Then as $M\rightarrow\infty$, the random variable $\sqrt M(\Bar{\bm{X}}-\mu)$ converges in distribution to a normal $\mathcal{N}(0,\sigma^2)$.

We start with the Mean Value Theorem for the function $L_M'$, the derivative of $L_M$ (continuous by assumption), on the interval $[\Theta_{\mathrm{MLE}}, \theta_0]$: 
\begin{align*}\label{eq:47}
    &0=L_M'(\Theta_{\mathrm{MLE}})=L_M'(\theta_0)+L_M''(\theta_1)(\theta_0-\Theta_{\mathrm{MLE}})\nonumber\\
    &\implies \theta_0-\Theta_{\mathrm{MLE}} = -\frac{L_M'(\theta_0)}{L_M''(\theta_1)}\nonumber\\
    &\implies \sqrt{M}(\theta_0-\Theta_{\mathrm{MLE}}) = -\sqrt{M}\frac{L_M'(\theta_0)}{L_M''(\theta_1)}\numberthis
\end{align*}
for some $\theta_1\in[\Theta_{\mathrm{MLE}}, \theta_0]$. We analyze the numerator and denominator respectively. The numerator
\begin{align*} \label{eq:48}
    L_M'(\theta_0)=&\frac{1}{M}\sum_{i=1}^M \left(\log P(X_i|\theta_0)\right)'\nonumber\\
    =&\frac{1}{M}\sum_{i=1}^M \left(\log P(X_i|\theta_0)\right)'-L'(\theta_0)\nonumber\\
    =&\frac{1}{M}\sum_{i=1}^M \left(\log P(X_i|\theta_0)\right)'-\langle \left(\log P(\bm{X}|\theta_0)\right)'\rangle_{\theta_0}\nonumber\\
    =&\frac{1}{M}\left(\sum_{i=1}^M \log P(X_i|\theta_0)\right)'-\langle \left(\log P(X_i|\theta_0)\right)'\rangle_{\theta_0}\numberthis
\end{align*}
where the last equality is obtained from the linearity of expected value and derivative. By the CLT, the distribution of $ \sqrt{M} L_M'(\theta_0)$ converges to
\begin{equation} \label{eq:49}
    \mathcal{N}\bigg(0,\mathrm{Var}_{\theta_0} (\log P(X_i|\theta_0))'\bigg)\numberthis
\end{equation} where the variance
\begin{align*} \label{eq:50}
    \mathrm{Var}_{\theta_0} &(\log P(X_i|\theta_0))'=\langle[(\log P(X_i|\theta_0))']^2 \rangle_{\theta_0}-\langle (\log P(X_i|\theta_0)'\rangle^2_{\theta_0}\nonumber\\
    &=\sum_{x\in X_1} P(x|\theta_0) \left(\frac{P'(x|\theta_0)}{P(x|\theta_0)}\right)^2-(L'(\theta_0))^2\nonumber\\
    &=\mathrm{CFI}_{\theta_0}\numberthis
\end{align*}
by the definition of CFI and that $\theta_0$ maximizes $L(\theta)$.
By the consistency property, $\Theta_{\mathrm{MLE}}$ converges to $\theta_0$, and thus $\theta_1$ converges to $\theta_0$. The denominator
\begin{align*} \label{eq:51}
    L_M''(\theta_1)&\rightarrow L_M''(\theta_0)=\frac{1}{M}\sum_{i=1}^M [\log P(X_i|\theta_0)]''\rightarrow \langle [\log P(X_1|\theta_0)]''\rangle_{\theta_0}\numberthis
\end{align*}
by the WLLN. We further show that Eq.~(\ref{eq:51}) is in fact the additive inverse of CFI: 
\begin{align*} \label{eq:52}
    \langle [\log P(X_1&|\theta_0)]''\rangle_{\theta_0} =\bigg\langle \frac{\partial^2}{\partial\theta^2}\log P(X_1|\theta_0)\bigg\rangle_{\theta_0}\nonumber\\
    &=\sum_{x\in X_1} [\log P(x|\theta_0)]'' P(x|\theta_0)\nonumber\\
    &=\sum_{x\in X_1} \left(\frac{P''(x|\theta_0)}{P(x|\theta_0)}-\left(\frac{P'(x|\theta_0)}{P(x|\theta_0)}\right)^2 \right)P(x|\theta_0)\nonumber\\
    &=\sum_{x\in X_1}P''(x|\theta_0)-\sum_{x\in X_1}\frac{(P'(x|\theta_0))^2}{P(x|\theta_0)}\nonumber\\
    &=0-\mathrm{CFI}_{\theta_0}=-\mathrm{CFI}_{\theta_0}.\numberthis
\end{align*}
Finally, Eq.~(\ref{eq:47}) becomes
\begin{align*} \label{eq:53}
    \sqrt{M}(\theta_0-\Theta_{\mathrm{MLE}})&\overset{p}{\to}\mathcal{N}\left(0,\frac{\mathrm{CFI}_{\theta_0}}{\mathrm{CFI}_{\theta_0}^2}\right)=\mathcal{N}\left(0,\frac{1}{\mathrm{CFI}_{\theta_0}}\right)\nonumber\\
    \implies \Theta_{\mathrm{MLE}} &\overset{p}{\to}\mathcal{N}\left(\theta_0,\frac{1}{M\cdot\mathrm{CFI}_{\theta_0}}\right)\numberthis
\end{align*}
Thus, the MLE is asymptotically unbiased and saturates the Cram\'{e}r-Rao bound.

\subsection*{Master equation for a non-Markovian environment}
To simulate the performance of our optimized states during the Ramsey measurement with non-Markovian noise, we use a time-local master equation given by [\onlinecite{smirne2016ultimate}]. A brief summary of the derivation is given below. 

\begin{enumerate}
    \item Let $\mathcal{L}(\mathbb{C}^d)$ be the Hilbert space of linear operators acting on $\mathbb{C}^d$, where the inner product is defined as $\langle\sigma,\tau\rangle=\text{Tr}(\sigma^\dagger\tau)$ (the Hilbert-Schmidt inner product).
    
    \item Let $\mathcal{LL}(\mathbb{C}^d)$ be the Hilbert space of linear operators acting on $\mathcal{L}(\mathbb{C}^d)$ which has dimension $d^2\times d^2$. Let $\{l_i\}_{i=1,...,d^2}$ be an orthonormal basis of $\mathcal{LL}(\mathbb{C}^d)$. Then the action of $\Lambda\in\mathcal{LL}(\mathbb{C}^d)$ on $\tau\in\mathcal{L}(\mathbb{C}^d)$ can be expressed as 
    \begin{align} \label{eq:Lambda on tau}
    \Lambda[\tau]=\sum_{ij=1}^{d^2}\langle l_i,\Lambda[l_j]\rangle\langle l_j,\tau\rangle l_i.\numberthis
    \end{align} 
    Thus, $\Lambda$ has a unique correspondence with the matrix $\mathsf{\Lambda}$ with entries
    \begin{align} \label{eq:matrix rep of Lambda}
      \mathsf{\Lambda}_{ij}\equiv\langle l_i,\Lambda[l_j]\rangle.\numberthis  
    \end{align}
    
    \item $\Lambda \in \mathcal{LL}(\mathbb{C}^d)$ is trace- and hermicity-preserving if and only if its matrix representation $\mathsf{\Lambda}$ can be written as 
    \begin{align} \label{eq:trace-herm-pres-map}
    \begin{pmatrix}
    1 & \bm{0}\\
    \mathsf{\bm{m}} & \mathsf{\bm{M}}
    \end{pmatrix},\numberthis
    \end{align} where $\bm{0}$ is the zero row vector of length $d^2-1$, $\mathsf{\bm{m}}$ is a real column vector of length $d^2-1$, and $\mathsf{\bm{M}}$ is a  $(d^2-1)(d^2-1)$ real matrix.
    
    \item  For a single qubit, any operator $\rho$ on $\mathbb{C}^2$ can be written as \begin{align} \label{eq:58}
        \rho=\frac{1}{2}(\bm{I}+\bm{v}\cdot\bm{\sigma})\numberthis
    \end{align} where $\bm{v}$ is a three-dimensional real vector and $\bm{\sigma}$ is the vector of Pauli matrices. Then a map $\Lambda$ whose matrix representation is given by Eq. (\ref{eq:trace-herm-pres-map}) acting on $\rho$ gives 
    \begin{align} \label{eq:Lambda on rho}
    \Lambda [\rho]=\frac{1}{2}(\bm{I}+(\mathsf{\bm{m}}+\mathsf{\bm{M}}\bm{v})\cdot\bm{\sigma}).\numberthis
    \end{align}
    
    \item The noisy evolution of a state $\rho$ is described by 
    \begin{align} \label{eq:evolution of rho}
    \rho(t)=\Lambda(t)[\rho(0)].\numberthis
    \end{align}
    The time local master equation satisfies 
    \begin{align} \label{eq:tlme1}
    \frac{d}{dt}\rho(t)=\Xi(t)[\rho(t)].\numberthis
    \end{align}
    So  
    \begin{align} \label{eq:tlme2}
    \Xi(t)=\frac{d\Lambda(t)}{dt}\circ\Lambda(t)^{-1}\numberthis
    \end{align} 
    with the corresponding matrix representation 
    \begin{align} \label{eq:tlme matrix rep}
    \mathsf{\Xi}(t)=\frac{d\mathsf{\Lambda}(t)}{dt}\mathsf{\Lambda}(t)^{-1}.\numberthis
    \end{align} 
    
    \item Consider the evolution of one qubit described by $\Lambda(t)=\mathcal{U}(t)\circ \Gamma(t)$ . $\mathcal{U}(t)$ is defined as 
    \begin{align} \label{eq: def of u}
    \mathcal{U}(t)[\rho(0)]\equiv U(t)\rho(0) U^\dagger(t) \numberthis
    \end{align}
    where $U(t)=e^{-i \frac{\omega t}{2} \sigma_z}$ represents the signal accumulation. By Eq. (\ref{eq:Lambda on tau}) and Eq. (\ref{eq: def of u}), the matrix representation of $\mathcal{U}(t)$ is
    \begin{align} \label{eq: matrix rep of U(t)} \mathsf{U}(t)=
    \begin{pmatrix}
    1 & 0 & 0 & 0\\
    0 & \cos{\omega t} & -\sin{\omega t} & 0\\
    0 & \sin{\omega t} & \cos{\omega t} & 0\\
    0 & 0 & 0 & 1
    \end{pmatrix} \numberthis.
    \end{align} 
    
    $\Gamma(t)$ represents the noise which is trace- and hermicity- preserving, i.e., has the form in Eq. (\ref{eq:trace-herm-pres-map}).
    
    \item Solving the commutation relation that gives phase covariant qubit map [\onlinecite{smirne2016ultimate},\onlinecite{holevo1993note}]
    \begin{align} \label{commute relation}
        [\mathcal{U}(t),\Gamma(t)]=0\iff[\mathsf{U}(t),\mathsf{\Gamma}(t)]=0,\numberthis
    \end{align} we obtain the matrix representation of $\Lambda(t)$:
    \begin{align} \label{eq:61}
        \mathsf{\Lambda}(t)=
        \begin{pmatrix}
        1 & 0 & 0 & 0\\
        0 & \eta_\perp(t)\cos{\omega t} & -\eta_\perp(t)\sin{\omega t} & 0\\
        0 & \eta_\perp(t)\sin{\omega t} & \eta_\perp(t)\cos{\omega t} & 0\\
        \kappa(t) & 0 & 0 & \eta_\parallel(t)
        \end{pmatrix},\numberthis
    \end{align} where $\mathsf{\bm{m}}=(0,0,\kappa(t))^T$ describes a translation along the $z$-axis, and $\mathsf{\bm{M}}= \begin{pmatrix}
        \eta_\perp(t)\cos{\omega t} & -\eta_\perp(t)\sin{\omega t} & 0\\
        \eta_\perp(t)\sin{\omega t} & \eta_\perp(t)\cos{\omega t} & 0\\
        0 & 0 & \eta_\parallel(t)
        \end{pmatrix}$ describes a rotation along the $z$-axis and a contraction characterized by $\eta_\perp$ and $\eta_\parallel$. 
        
    \item By Eq.~(\ref{eq:tlme matrix rep}), we obtain the time-local master equation for a single qubit:
    \begin{align*} \label{eq:62}
        \Xi(t)[\rho(t)]=& -\frac{i}{2}\omega[\sigma_z,\rho(t)]\nonumber\\
        &+ \gamma_+(t)(\sigma_+\rho(t)\sigma_- - \frac{1}{2}\{\sigma_-\sigma_+,\rho(t)\})\nonumber\\
        &+ \gamma_-(t)(\sigma_-\rho(t)\sigma_+ - \frac{1}{2}\{\sigma_+\sigma_-,\rho(t)\})\nonumber\\
        &+ \gamma_z(t)(\sigma_z\rho(t)\sigma_z-\rho(t)),\numberthis
    \end{align*}
    where
    \begin{align} \label{eq:63}
        \gamma_+(t)&=\frac{1}{2}\left(\kappa^{\mathrm{'}}(t)-\frac{\eta^{\mathrm{'}}_{\mathrm{\parallel}}(t)}{\eta_\parallel(t)}(\kappa(t)+1)\right)\numberthis,\\
        \gamma_-(t)&=-\frac{1}{2}\left(\kappa^{\mathrm{'}}(t)+\frac{\eta^{\mathrm{'}}_{\mathrm{\parallel}}(t)}{\eta_\parallel(t)}(1-\kappa(t))\right)\numberthis,\\
        \gamma_z(t)&=\frac{1}{4}\left(
        \frac{\eta^{\mathrm{'}}_{\mathrm{\parallel}}(t)}{\eta_\parallel(t)}-2\frac{\eta^{\mathrm{'}}_{\mathrm{\perp}}(t)}{\eta_\perp(t)}
        \right)\numberthis.
    \end{align}
    \end{enumerate}

Considering only $T_2$ noise, $\gamma_-(t)=\gamma_+(t)=0$, $\eta_\parallel$ is constant, and 
\begin{align} \label{eq:64}
    \eta_\perp(t)=e^{-(\frac{t}{T_2})^\nu},\numberthis
\end{align}
where $\nu$ is the stretch character which equals $1$ for Markovian noise. Then 
\begin{align} \label{eq:65}
    \gamma_z(t)=\frac{\nu}{2}\frac{t^{\nu-1}}{T_2^\nu}.\numberthis
\end{align}

We further need to express $\Xi(t)$ as a superoperator acting on the vectorization of $\rho(t)$. Defining the vectorization of a matrix as the map
\begin{align} \label{eq:66}
    \rho=\sum_{i,j} \rho_{ij}\ket{i}\bra{j}\mapsto\ket{\rho}=\sum_{i,j} \rho_{ij}\ket{j}\otimes\ket{i}.\numberthis
\end{align}
Define the left and right multiplication superoperators by $\mathcal{L}(A)[\rho]=A\rho$ and $\mathcal{R}(A)[\rho]=\rho A$ so that $[A,\rho]=\mathcal{L}(A)[\rho]-\mathcal{R}(A)[\rho]$. By this definition, we can calculate the matrix representation $\mathcal{L}(A)=I\otimes A$ and $\mathcal{R}(A)=A^T\otimes I$.
Using the superoperator notation, we can express $\Xi(t)$ as 
\begin{align*} \label{eq:67}
    \Xi(t)=&-\frac{i}{2}\omega(I\otimes\sigma_z - \sigma_z\otimes I)+\gamma_z(t)(\sigma_z\otimes\sigma_z-I\otimes I).\numberthis
\end{align*}
With this expression, we numerically simulate the evolution of our entangled states under non-Markovian noise by using the Time-Dependent Master Equation Solver in QuTip [\onlinecite{qutip}].

\subsection*{Performance of metrological states in a non-Markovian environment}
To calculate the derivative of probability with respect to $\omega$ in the calculation for $\mathrm{CFI}_\omega$, we use a method similar to parameter shift that utilizes the property that the signal accumulation operator ($\mathcal{U}(\omega)=e^{-i\omega t J_z}$) and the noisy operator commutes. In the following notation, 
\begin{enumerate}
    \item $z$ represents a multi-qubit state in the $z$ basis;
    \item $\mathcal{U}(\omega)$ is the effective signal accumulation operator: $\mathcal{U}(\omega)=e^{-i\omega t J_y}$;
    \item $\rho$ is the state density matrix of our optimized state after the noisy evolution without signal for some Ramsey time and a $\frac{\pi}{2}$ pulse along the $x$ direction (here we switch the order of the signal accumulation and the second pulse of the Ramsey protocol [\onlinecite{strobel2014fisher}]);
    \item $P(z|\omega)$ is the probability of measuring the state $z$ with our rotated optimized state after the noisy evolution and signal accumulation.
\end{enumerate}
Then 
\begin{align*} \label{eq:69}
    \frac{\partial}{\partial \omega} P(z|\omega) \Bigr\rvert_{\omega \rightarrow 0}&= \frac{\partial}{\partial \omega} \mathrm{Tr}[\ket{z}\bra{z}\mathcal{U}(\omega)\rho\mathcal{U}^\dagger(\omega)]\Bigr\rvert_{\omega \rightarrow 0}\\
    &= \frac{\partial}{\partial \omega}\mathrm{Tr}[\mathcal{U}^\dagger(\omega)\ket{z}\bra{z}\mathcal{U}(\omega)\rho]\Bigr\rvert_{\omega \rightarrow 0}\\
    &= \mathrm{Tr}[\frac{\partial}{\partial \omega}\mathcal{U}^\dagger(\omega)\ket{z}\bra{z}\mathcal{U}(\omega)\rho]\Bigr\rvert_{\omega \rightarrow 0}+\mathrm{Tr}[\mathcal{U}^\dagger(\omega)\ket{z}\bra{z}\frac{\partial}{\partial \omega}\mathcal{U}(\omega)\rho]\Bigr\rvert_{\omega \rightarrow 0}\\
    &=\numberthis{i t \mathrm{Tr}[(J_y \ket{z}\bra{z}- \ket{z}\bra{z}J_y)\rho]}.
\end{align*}
From Eq.~(\ref{eq:37}), since $T_\text{tot}$ and $\delta\omega^2$ are constants, SNR is proportional to $\sqrt{\frac{\mathrm{CFI}_\omega}{t_{\text{R}}}}$. Thus we choose $\frac{\mathrm{CFI}_\omega}{t_{\text{R}}}$ as the result we show in Fig.4(d) in the main text. 

\subsection*{Time Overhead}

\begin{figure}[h]
\includegraphics[width=80mm]{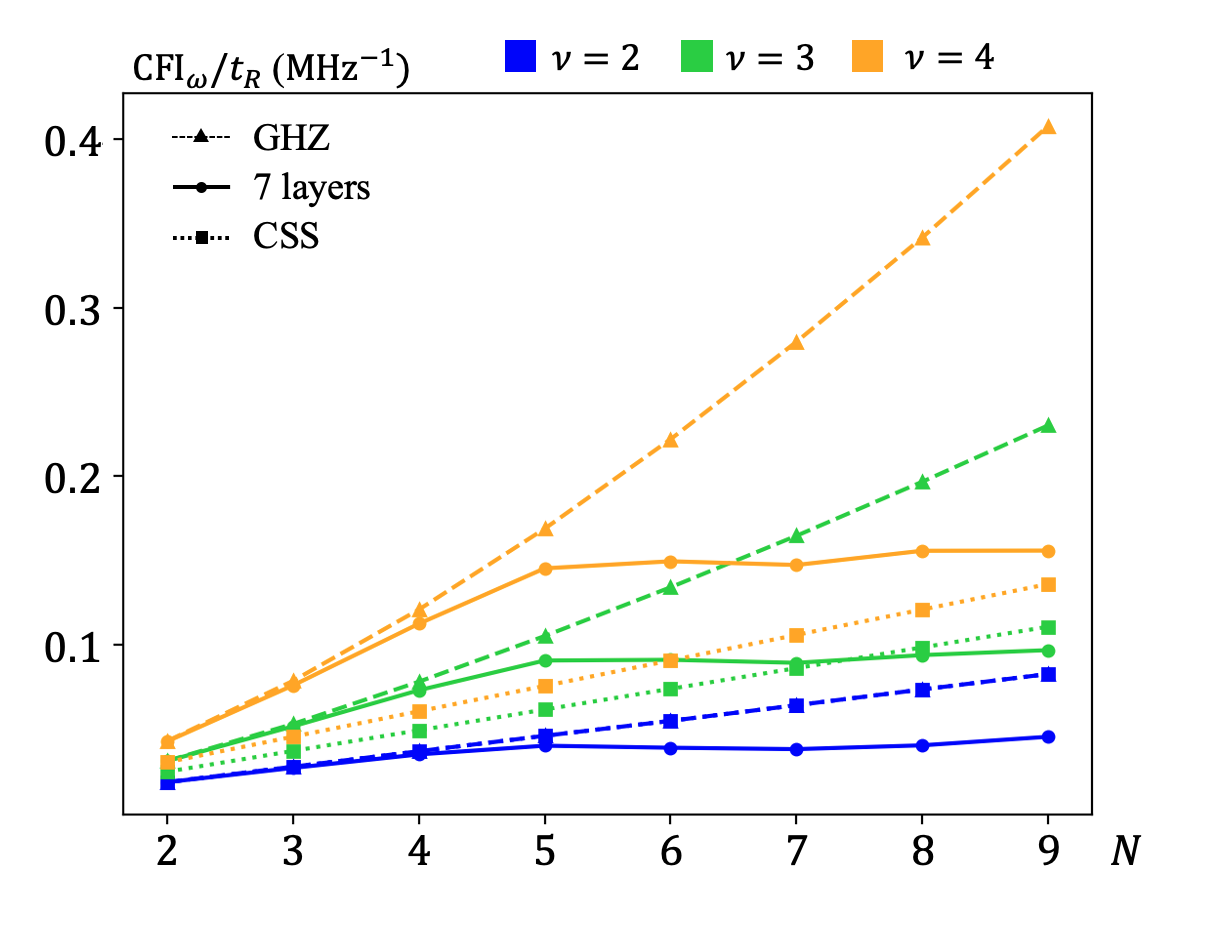}
\caption{\label{fig:time_overhead}50 cases average sensing performance of the optimized states when using 7-layer circuit on 3D random spin configuration.}
\end{figure}

In experiments, the time overhead, including the state preparation and readout time, reduces the repetition number of the sensing sequence and thus decreases the sensitivity. If we consider a nonzero time overhead, i.e., $M=\frac{T_\text{tot}}{t_\text{R}+t_\text{oh}}$, the expression for $\mathrm{SNR}^2$ for an uncorrelated spin state becomes

\begin{align} \label{eq:70}
\mathrm{SNR}^2 = \frac{T_\text{tot} t_\text{R}^2 \cos^2(\omega t_\text{R}) \delta \omega^2}{(t_\text{R}+t_\text{oh})\left(e^{2\left(\frac{t_\text{R}}{T_2}\right)^\nu} - \sin^2(\omega t_\text{R})\right)}.\numberthis
\end{align}
If $t_\text{oh}>>t_\text{R}$, we ignore the term $t_\text{R}$ in the denominator and 

\begin{align} \label{eq:71}
    \mathrm{SNR}^2 \propto \frac{t_\text{R}^2}{e^{2\left(\frac{t_\text{R}}{T_2}\right)^\nu}}.\numberthis
\end{align} 

Taking the derivative of Eq.~(\ref{eq:71}) with respect to $t_\text{R}$ gives us the best $t_\text{R}$ if the time overhead is significantly larger:
\begin{align}\label{eq:72}
    t_\text{R} = \frac{T_2}{\nu^\frac{1}{\nu}}.\numberthis
\end{align}
Similarly, the same calculations for a GHZ state where the decay term in Eq.~(\ref{eq:70}) becomes $e^{2n\left(\frac{t_\text{R}}{T_2}\right)^\nu}$ show that the best Ramsey sensing time is 
\begin{align}\label{eq:73}
    t_\text{R} = \frac{T_2}{(n\nu)^\frac{2}{\nu}}.\numberthis
\end{align}
Plugging Eq.~(\ref{eq:72}) and Eq.~(\ref{eq:73}) into Eq.~(\ref{eq:71}), we find that the ratio of the $\mathrm{SNR}^2$ of a GHZ state to that of an uncorrelated spin state is $n^{1-\frac{2}{\nu}}$. Thus, only when 
\begin{align}\label{eq:mu2}
    \nu>2
    \numberthis
\end{align}
do GHZ states provide an advantage in SNR over uncorrelated spin states when $t_\text{oh}>>t_\text{R}$. We compare the SNR of the states generated by the optimizer with that of the CSS and GHZ states when $\nu=2,3,4$. Fig.~\ref{fig:time_overhead} shows that when we assume a long time overhead, the generated entangled states are less sensitive than CSS when $\nu=2$ and $\nu=3$ for large spin numbers.

\subsection*{State preparation time comparing to adiabatic method}
State preparation time is one of the major components of the time overhead in the generalized Ramsey sensing sequence which influences the sensitivity. The state preparation time of the variational method depends on the circuit layer number $m$, system size $N$ and is proportional to the inverse of average interaction strength $1/\bar{f}_\text{dd}$. The adiabatic method [\onlinecite{cappellaro2009quantum}] is an alternative approach to generate entangled states for quantum metrology in dipolar-interacting spin systems by only using single-qubit rotations (global pulses). 

To compare the performance of our variational method with the adiabatic method, we derive the relation between the squeezing parameter (Wineland parameter [\onlinecite{wineland1994squeezed}]) and CFI. Without loss of generality, we consider a SSS with collective spin direction $+x$ and is squeezed along the $y$-axis (such as the 3rd Wigner distribution shown in Fig.~\ref{fig:Optimization_with_RF}(b)). In this case, the squeezing parameter is
\begin{align}\label{eq:74}
    \xi^2 = N\frac{(\Delta J_y)^2}{|\langle J_x\rangle|^2},
    \numberthis
\end{align}
where $(\Delta O)^2  \equiv \langle O^2\rangle - \langle O\rangle^2$ and $N$ is the number of spins.
According to the uncertainty principle,

\begin{align}\label{eq:75}
    (\Delta J_y)^2(\Delta J_z)^2 \geq \frac{1}{4}|\langle J_x\rangle|^2.
    \numberthis
\end{align}
The relation between the squeezing parameter and total spin angular momentum uncertainty projection in $z$-direction is
\begin{align}\label{eq:76}
    4(\Delta J_z)^2 \geq \frac{|\langle J_x\rangle|^2}{(\Delta J_y)^2} = N/\xi^2.
    \numberthis
\end{align}
It's been proven that for a pure Gaussian state, the quantum Fisher information (QFI) is directly related to the variance of the projected spin angular momentum [\onlinecite{braunstein1994statistical},\onlinecite{pezze2009entanglement},\onlinecite{hyllus2010not}]:
\begin{align}\label{eq:77}
    \text{QFI} = 4(\Delta J_z)^2.
    \numberthis
\end{align}
Combining Eq.~(\ref{eq:76}) and Eq.~(\ref{eq:77}), we obtain the relation between CFI and squeezing parameter of a SSS:
\begin{align}\label{eq:cfi_qfi_sp}
    \text{CFI}\leq\text{QFI} \geq N/\xi^2.
    \numberthis
\end{align}
The first inequality in Eq.~(\ref{eq:cfi_qfi_sp}) is saturated by measuring the SSS along the direction where it is squeezed ($y$-axis, or equivalently measuring it in $z$-basis after applying a $R_x(\frac{\pi}{2})$ pulse [\onlinecite{pedrozo2020entanglement}]). The second inequality originates from the uncertainty principle (Eq.~(\ref{eq:75})). Since the optimal SSS saturate the Heisenberg uncertainty relation [\onlinecite{pezze2018quantum}] and the SSS generated by the adiabatic method [\onlinecite{cappellaro2009quantum}] belongs to these states, we obtain the relation between the squeezing parameter and CFI 
\begin{align}\label{eq:78}
    \text{CFI} = N/\xi^2.
    \numberthis
\end{align}

Based on the data shown in Fig.3 from ref.~(\onlinecite{cappellaro2009quantum}), it takes about $200\mu\text{s}$ for the adiabatic method to prepare an 8-spin SSS with $\xi^2 = 0.4$ which corresponds to $\text{CFI} = 20$. The 2D spin density $8/(30\text{nm}\times30\text{nm})$ corresponds to $\bar{f}_{\text{dd}} = 43.5\text{kHz}$. According to Fig.3(d) in the main text, the variational method is able to prepare an 8-spin entangled state with $\text{CFI} \approx 20$ by a 4-layer circuit with $\bar{f}_{\text{dd}}T = 0.8$. Plugging in the same average nearest neighbor dipolar interaction strength $\bar{f}_{\text{dd}}$, we finally calculated the state preparation time of the variational method is $T = 18.4\mu\text{s}$, which is about 11 times faster than the adiabatic method under the same condition.

\section*{Controllability}

Since all the black-box optimization algorithms cannot ensure that the optimized result is the global maximum/minimum point of in the parameter space, it is sill an open question that if the variational method is able to find the 'best' metrological state for a given spin configuration or not. In this section, we're interested in the theoretically achievable controllability of dipolar interacting spin systems. The question is, given any (possibly infinite) arbitrary sequence of evolution under each Hamiltonian governing the dynamics of our system, can we drive any arbitrary unitary operator? Quantum control systems of the general form
\begin{equation}
    H(t) = H_0 + \sum_{k=1}^K u_k(t)H_k, \label{eq:control-system}\numberthis
\end{equation}
governed by the Schr\"odinger equation, $i\frac{d}{dt}\ket{\psi(t)}=H(t)\ket{\psi(t)},$ have been studied extensively [\onlinecite{d2021introduction},\onlinecite{schirmer2001complete}]. $H_0$ is the unperturbed or free evolution Hamiltonian, $H_k$ are the control interactions, and $u_k(t)$ are the piecewise continuous control fields. There are several distinct but related notions of controllability that have different conditions for `full' controllability. The notion of `operator' or `complete' controllability is the strictest condition and is defined as above. For generic interacting spin systems, all of these notions are equivalent. Complete controllability is equivalent to universal quantum computation (UQC) in quantum information processing (QIP) [\onlinecite{wang2016subspace},\onlinecite{ramakrishna1996relation}].

\subsection*{Controllability Test}

The way we investigate the controllability of a generic system (\ref{eq:control-system}) is by examining the so-called `dynamical Lie algebra' $\mathcal{L}_0\subseteq u(\mathcal{N})$ or $su(\mathcal{N})$ generated by the operators $\{-iH_0,-iH_1,\ldots,-iH_K\},$ which are represented by $\mathcal{N}\times \mathcal{N}$ matrices in a basis we choose [\onlinecite{d2021introduction},\onlinecite{schirmer2001complete}].

A quantum system of the form (\ref{eq:control-system}) is completely controllable if either $\mathcal{L}_0\cong{}u(\mathcal{N})$ or $\mathcal{L}_0\cong{}su(\mathcal{N})$ [\onlinecite{d2021introduction}], where $u(\mathcal{N})$ is the unitary Lie algebra represented by the set of skew-Hermitian $\mathcal{N}\times \mathcal{N}$ matrices and $su(\mathcal{N})$ is the special unitary Lie algebra represented by the same set of matrices with the extra condition that they are traceless. Note that $\dim{u(\mathcal{N})}=\mathcal{N}^2$ and $\dim{su(\mathcal{N})}=\mathcal{N}^2-1$, and the difference of $1$ comes from counting identity operation ($I$) as a dimension or not. We must find a basis for $\mathcal{L}_0$ by iteratively taking the Lie bracket $[\cdot,\cdot]$ of $H_0,H_1,\ldots,H_K$ until we have a set of $\dim{\mathcal{L}_0}$ linearly independent matrices, where the Lie bracket is the commutator $[A,B]=AB-BA$ for matrices $A$ and $B$. Ref.[\onlinecite{d2021introduction}] and ref.[\onlinecite{schirmer2001complete}] present an algorithm for generating this basis. Thus, if $\dim{\mathcal{L}_0}=\mathcal{N}^2$ or $\mathcal{N}^2-1$ we can say that the system is completely controllable. Note that for generic spin systems $\mathcal{N}=2^N$ for $N$ spins.

\begin{table}[]
\begin{tabular}{l}
\hline
\multicolumn{1}{c}{\textbf{Algorithm.} Generating $\mathcal{L}_0$ and finding $\dim{\mathcal{L}_0}$.} \\ \hline
\textit{Input:} Hamiltonians $I\equiv \{H_0,H_1,\ldots,H_K\}$\\
1. $B\equiv$ maximal linearly independent subset of $I$\\
2. $r\equiv \abs{B}$\\
3. If $r=N^2$ then $O\equiv B$ else $O\equiv \{\}$\\
4. If $r=N^2$ or $\abs{B}=0$ then terminate\\
5. $C\equiv[O,B]\cup[B,B]$, where \\
$\,\quad[S_1,S_2]\equiv\{[s_1,s_2]\,|\,s_1\in S_1,s_2\in S_2\}$\\
6. $O=O\cup B$\\
7. $B=$ maximal linearly independent extension of $O$ with$\,\,\,\,\,$\\
$\quad$ elements from $C$\\
8. $r=r+\abs{B}$; Go to 4\\
\textit{Output:} basis $O$ of $\mathcal{L}_0$ and $\dim{\mathcal{L}_0}=r$\\ \hline
\end{tabular}
\caption{Implementation of [\onlinecite{schirmer2001complete}]'s algorithm with a few physically motivated modifications. Note $\abs{S}$ indicates the cardinality of set $S$.}
\label{tab:algorithm}
\end{table}

\subsection*{Controllability of Dipolar Interacting Spin Systems}
We write our system in the form (\ref{eq:control-system}) by defining the free evolution Hamiltonian to be the dipolar interaction $H_{\text{dd}}$ and two control interactions $J_x$ and $J_y$, as these operators are generators of rotation, with respective independent control fields $\theta_x(t)$ and $\theta_y(t)$:
\begin{equation}
    H(t)=H_\text{dd}+\theta_x(t)J_x + \theta_y(t)J_y. \label{eq:dipsyscontrollable}\numberthis
\end{equation}
Ref.[\onlinecite{schirmer2001complete},\onlinecite{polack2009uncontrollable}] demonstrate that we cannot achieve complete controllability with global controls due to inherent symmetries, so we know that $\dim{\mathcal{L}_0}<4^N-1$.

However, complete controllability is a rather strict condition. Not being able to drive any arbitrary unitary does not mean we cannot drive unitaries that produce metrological states.

In fact, ref.[\onlinecite{chen2017preparing}] demonstrate for a long-range Ising spin model (all-to-all interactions) with global controls that metrological states, such as the GHZ and W states are reachable. Ref.[\onlinecite{albertini2018controllability}] extend their result for symmetric Ising spin networks with global controls and demonstrate that one can reach any state that preserves spin permutation invariance. This is known as subspace controllability. The dimension of their dynamical Lie algebra, $\mathcal{L}^\text{Ising}\equiv\mathcal{L}^{\text{PI}}\cap su(2^N)$, is shown to be $\binom{N+3}{N}-1$. This is relevant to our system because [\onlinecite{albertini2021subspace}] show that if we replace the Ising interaction with a more general two body interaction---which includes $H_{\text{dd}}$---the dimension of the dynamical Lie algebra is necessarily greater than or equal to that of the symmetric Ising case, and it is therefore subspace controllable. This means that we can write $\binom{N+3}{N}-1\leq\dim{\mathcal{L}_0}<4^N - 1$ and say that $\mathcal{L}_0$ is subspace controllable but not completely controllable. Therefore, we can achieve arbitrary permutation invariant states, including metrological states such as a GHZ state.

\begin{table}
\begin{ruledtabular}
\begin{tabular}{c|lllll}
Lie algebra dimension & $N=2$ & $N=3$ & $N=4$ & $N=5$  \\
Completely controllable: $4^N$ (or $4^N-1$) & 16 & 64 & 256 & 1024\\
$H_\text{dd}$ & 9 & 39 & 225\\
Symmetric Ising: $\binom{N+3}{N}-1$ & 9 & 19 & 34 & 55
\end{tabular}
\end{ruledtabular}

\caption{Lie algebra dimensions for the complete controllable system, dipolar interacting system and symmetric Ising system (lower bound for subspace controllability). Dipolar interacting spin systems' $\dim{\mathcal{L}_0}$ is calculated using an implementation of [\onlinecite{schirmer2001complete}]'s algorithm, and is necessarily bounded by the complete and subspace controllability dimensions. Lie algebra dimensions for dipolar interacting systems are only calculated up to $N=4$ due to stability issues stemming from numerical errors in how matrix rank is calculated.}
\label{table:Lie_dimension}
\end{table}

\subsection*{Finding Reachable States}
$\mathcal{L}_0$ is associated with a Lie group $e^{\mathcal{L}_0}$ by the Lie group–Lie algebra correspondence [\onlinecite{d2021introduction}]. The Lie algebra $u(\mathcal{N})$ corresponds to the Lie group $U(\mathcal{N})$, and $su(\mathcal{N})$ corresponds to $SU(\mathcal{N})$. We can define $\mathcal{R}\equiv e^{\mathcal{L}_0}$ as the reachable set of unitaries we can drive under $\{H_k\}_{k=0,\ldots,K}$, and so starting from an initial state $\ket{\psi_0}, \mathcal{R}_{\ket{\psi_0}}$ is the set of states we can reach.

As demonstrated in the previous section, our dynamical Lie algebra is a superset of $\mathcal{L}^\text{Ising}$ and a strict subset of $su(2^N)$, so we can write $e^{\mathcal{L}^\text{Ising}}\subseteq e^{\mathcal{L}_0}\subset SU(2^N)$. Because $\ket{\text{GHZ}}\in \mathcal{R}_{\ket{0}^{\otimes N}}^\text{Ising}$ we can write $\ket{\text{GHZ}}\in \mathcal{R}_{\ket{0}^{\otimes N}}^\text{dipolar}$. In fact, this is true for any permutation invariant state, which includes all metrological states we're interested in.

While we know that metrological states are in the reachable set, determining the parameters that drive the unitaries to produce those states is a highly convex optimization problem equivalent to our variational circuit, using state fidelity between the ideal state and the current state instead of CFI as the cost function. That is we optimize the output unitary of the variational circuit,
\begin{equation}
    \mathcal{S}(\bm{\theta})=e^{-i\frac{\pi}{2}J_y}\prod_{i=1}^m e^{-i\tau_iH_\text{dd}} e^{-i\vartheta_i J_x} e^{i\frac{\pi}{2}J_y} e^{-i\tau_i'H_\text{dd}} e^{-i\frac{\pi}{2}J_y},\label{eq:variationalcircuitproductform}\numberthis
\end{equation}
where $m$ is the (possibly infinite) number of layers, for state fidelity,
\begin{equation}
    \mathcal{F}(\ket{\text{GHZ}},\mathcal{S}(\bm{\theta})\ket{0}^{\otimes N})=\abs{\bra{\text{GHZ}}{\mathcal{S}(\bm{\theta})\ket{0}^{\otimes N}}}^2,\label{eq:fidelity}\numberthis
\end{equation}
for pure states. If there exists some $\bm{\theta}$ such that $\mathcal{F}(\ket{\text{GHZ}},\mathcal{S}(\bm{\theta})\ket{0}^{\otimes N})=1$, then we can say that $\ket{\text{GHZ}}\in \mathcal{R}_{\ket{0}^{\otimes N}}^\text{dipolar}$. From the previous section, we know such a $\bm{\theta}$ must exist, but it may be the case that $m\rightarrow \infty$, in which case it is not possible to find this exactly. This is the method employed in ref.[\onlinecite{chen2017preparing},\onlinecite{gao2013preparation}] to demonstrate the reachability of GHZ and W states for Ising spin models. Our variational circuit method represents an improvement in the efficiency of searching for such metrological states.


\end{appendix}

\end{document}